\documentclass[aps,pra,twocolumn,epsfig,floats]{revtex4}
\usepackage{amsmath}
\usepackage{color}
\usepackage{physics}
\usepackage{graphicx}
\usepackage{amsfonts}
\usepackage{mathrsfs}
\usepackage{amsmath}
\usepackage{amssymb}
\usepackage{float}
\usepackage[section]{placeins}
\usepackage[utf8x]{inputenc}
\usepackage{times}
\usepackage{hyperref}
\usepackage{multirow}
\hypersetup{
  colorlinks=true,
  citecolor=blue,
  linkcolor=blue,
  urlcolor=blue}

\begin{document} 

\title{ Formation of paired phases of bosons and their excitations in square lattice}
 
\author{Manali Malakar, Sudip Sinha, and S. Sinha}
\affiliation{Indian Institute of Science Education and
Research-Kolkata, Mohanpur, Nadia-741246, India}
 
\date{\today}
\begin{abstract}
We investigate the formation of paired states of bosons in an optical lattice, namely, pair superfluid (PSF) and pair supersolid (PSS) in the presence of pair hopping as well as the next nearest neighbor (NNN) interaction mimicking long-range forces. Both the zero and finite temperature phase diagrams are obtained using the cluster mean field theory, which includes the effect of correlations systematically. We also compute the low-energy excitations which capture the characteristic features of such paired states and their transitions. Apart from the gapless sound mode due to the PSF order, a gapped mode also appears in the PSF phase, similar to the Higgs mode of the usual atomic superfluid (ASF). The PSF to ASF transition exhibits an intriguing behavior due to the existence of a `tri-critical' point, where the nature of transition changes. As a consequence of the continuous PSF-ASF transition, the gapped mode of both the phases becomes gapless at the critical point. For sufficiently strong NNN interaction strength, a PSS phase appears with coexisting pair superfluidity and stripe density order. The softening of the roton mode as a precursor of density ordering and the appearance of a low-energy gapped mode serve as robust features related to the formation of the PSS phase. We also investigate the melting of PSF and PSS phases to normal fluid at finite temperatures, particularly the melting pathway of PSS which occurs in atleast two steps due to the coexisting orders. Finally, we discuss the possibility of emulating such exotic phases in the ongoing cold atom experiments.
\end{abstract}


\maketitle

\section{Introduction}
Advancement of quantum engineering techniques in ultracold atomic system has paved the way to realize various exotic phases of condensed matter physics \cite{Dalibard_review,Sanpera_review}. One of such example is supersolid phase with coexisting superfluidity and density order which has recently been observed experimentally in ultracold dipolar gas \cite{Modugno_SS_expt1,TPfau_SS_expt1,Ferlaino_SS_quench_expt1,
Ferlaino_SS_quench_expt2,Ferlaino_SS_2d_expt1,
Ferlaino_SS_quench_2d_expt}, although it has been predicted long back \cite{Penrose_BEC,Lifshitz,Chester,Leggett,Fisher,supersolids_review}. 
Similar to the Cooper pairs in superconductors, engineering the 
pairing mechanism between the bosons has attracted significant interest in recent years \cite{Daley_repulsive_bound_pairs,Bloch_second_order_tunneling,
Lukin_photon_bound_states}.
Pairing between two component or single component bosons can lead to the formation of fascinating pair superfluid (PSF) phase \cite{Rippol_PRA,PSF_square_lattice_QMC_Wessel,PSF_square_lattice_QMC_K_Ng,
Lewenstein_PRA,Yang_PRA,Daley_PRL,two_component_Kuklov_PRL,
two_component_Mathey_PRA,two_component_Stringari_PRA,
PSF_Quantum_Gutzwiller,two_layer_Luis,PSF_Bradraj,pair_tunneling_zhou,
Christophe_Mora_PRL,Tapan_Mishra_1,Tapan_Mishra_2,
Lewenstein_pair_hopping,Sandvik_pair_hopping_QMC}. Attractive interaction between bosons can generate paired states, however thermodynamic stability of such phase requires additional restriction on multiple occupancy of bosons at a single site which can be possible for strong three-body loss process \cite{PSF_square_lattice_QMC_Wessel,PSF_square_lattice_QMC_K_Ng,
Lewenstein_PRA,Yang_PRA,Daley_PRL}. Formation of the PSF phase due to the attractive interaction in single component bosons in an optical lattice in the presence of three-body constraint has been studied using the quantum Monte Carlo method where beyond mean field quantum effects play a significant role \cite{PSF_square_lattice_QMC_Wessel,PSF_square_lattice_QMC_K_Ng}. 
Theoretical studies reveal that such PSF phase can also be realised in two component bosons and in bilayer systems due to inter-species attraction \cite{two_component_Kuklov_PRL,
two_component_Mathey_PRA,two_component_Stringari_PRA,
PSF_Quantum_Gutzwiller,two_layer_Luis,PSF_Bradraj}. 
Alternatively, a recent proposal to engineer pair hopping of bosons in an optical lattice can also lead to the formation of bosonic paired states \cite{pair_tunneling_zhou}. Apart from the cold atom systems, circuit QED setups have also become a platform to manipulate such pair hopping of photons, which can lead to the possibility of creating interesting photonic cat states due to the pairing mechanism \cite{Christophe_Mora_PRL}. Recent theoretical studies have revealed the signature of pairing of photons in an atom-photon interacting system described by an array of Jayne Cummings model \cite{Sebastian_jc1}. 

Moreover, manipulating the pair hopping of bosons with long-range interaction not only gives rise to the PSF phase \cite{Lewenstein_pair_hopping}, but also can create an opportunity to realize the supersolid phase of paired bosons known as pair supersolid (PSS) \cite{Lewenstein_PSS}. Formation of PSS in triangular lattice due to the combined effect of pair hopping and lattice frustration has already been studied theoretically \cite{PSS_triangular_lattice_Zhang,CPB_Zhang}. Such unconventional quantum phases of bosons with coexisting pairing and density ordering deserve further analysis, particularly their regime of stability and low-lying excitations which are important for their characterization as well as their detection.

In this work, we primarily focus on the formation of pair superfluid and supersolid phases on a square lattice due to the presence of pair hopping and interactions beyond the on-site repulsion. We study the nature of transition from the atomic superfluid (ASF) to the PSF phase, which reveals a rather intriguing behavior. More importantly, the interplay between the pair hopping strength and the next nearest neighbor interaction leads to the formation of striped PSS phase via transition from the striped atomic supersolid phase (ASS). In addition, the low-energy collective modes can also characterize these phases as well as the transitions between them. Recent studies revealed that the softening of the `roton mode' in superfluids is a robust signature for the formation of supersolids \cite{Shlyapnikov_roton,Nozieres_roton}, which has also been observed in experiments \cite{Ferlaino_expt1_roton,Ferlaino_expt2_roton,Ferlaino_expt3_roton,
TPfau_expt1_roton,TPfau_expt2_roton,Esslinger_expt1_roton}. In the present context, it is also a pertinent issue to investigate whether such mode softening phenomena can also occur for the formation of PSS phase. Due to the presence of competing orders, the formation of correlated PSS phase of bosons is a delicate issue, since its stability may significantly be affected by the quantum fluctuations.
Moreover, the stability of such exotic paired states under thermal fluctuations, and their melting process deserves a careful analysis \cite{K_Ng_QMC_PSF}. Particularly, it is fascinating to understand the melting pathway of PSS phase due to the presence of two competing orders. The present study addresses the above issues related to the formation of paired phases of bosons and their characteristic features from low-lying excitations, which can be important for their detection.
To incorporate the effect of fluctuations beyond mean field in a systematic manner, we investigate such correlated phases by using the cluster mean field (CMF) technique \cite{Yamamoto_triangular_CMFT, Yamamoto_square_CMFT,Heydarinasab_CMFT,Demler_CMFT,Luhmann_CMFT,
Fazio_CMFT,Suthar_CMFT_1,Suthar_CMFT_2,Suthar_CMFT_3,Manali_CMFT_1,
Sayak_CMFT,Manali_CMFT_2} both at zero and finite temperatures.

The rest of the paper is organized as follows. In Sec.\ref{Model_and_Method}, we introduce the extended Bose Hubbard model and discuss the method to implement the mean field as well as cluster mean field technique. Next, we investigate the characteristic properties of the PSF and PSS phases at zero temperature in Sec.\ref{Formation_of_pair_SF_phase} and Sec.\ref{Formation_of_pair_SS_phase}, respectively. In Sec.\ref{finite_temperature_phase_diagram}, we obtain the finite temperature phase diagrams for the paired states within both the mean field as well as cluster mean field approach. Finally, we summarize and conclude our results in Sec.\ref{conclusion}.

\section{Model and method}
\label{Model_and_Method}
To investigate the different correlated phases of bosons due to the effect of long-range interaction and pairing, we consider an extended Bose-Hubbard model (EBHM) in the presence of pair hopping term, described by the following Hamiltonian $\hat{\mathcal{H}}=\hat{\mathcal{H}}_{\rm BHM} + \hat{\mathcal{H}}_{\rm p} + \hat{\mathcal{H}}_{\rm lr}$, with:
\begin{small}
\begin{subequations}
\begin{align}
\hat{\mathcal{H}}_{\rm BHM} =& -t\sum_{\langle ij \rangle}(\hat{a}^\dagger_{i}\hat{a}_{j}+\mathrm{h.c})\notag\\
&+ \frac{U}{2}\sum_{i}\hat{n}_{i}(\hat{n}_{i}-1)-\mu\sum_{i}\hat{n}_{i}\\
\hat{\mathcal{H}}_{\rm p} =&  -t_p\sum_{\langle ij \rangle}(\hat{a}^{\dagger2}_{i}\hat{a}^2_{j}+\mathrm{h.c})\\
\hat{\mathcal{H}}_{\rm lr} =&  V_{1}\sum_{\langle ij \rangle}\hat{n}_{i}\hat{n}_{j} + V_{2}\sum_{\langle\langle ij \rangle \rangle}\hat{n}_{i}\hat{n}_{j}
\end{align}
\label{hamiltonian}
\end{subequations}
\end{small}
where $\hat{a}_{i}^{\dagger}$ ($\hat{a}_{i}$) is the creation (annihilation) operator acting on the $i^{\rm{th}}$ lattice site. The first part $\hat{\mathcal{H}}_{\rm BHM}$ describes the usual BHM, where $t$ is the single particle hopping amplitude between the nearest neighbor sites, $U$ is the on-site repulsion, and $\mu$ denotes the chemical potential. The second part $\hat{\mathcal{H}}_{\rm p}$ describes the pair hopping of bosons with amplitude $t_{p}$ within the nearest neighbor (NN) sites. The last part $\hat{\mathcal{H}}_{\rm lr}$ describes the long-range interactions where for simplicity, we consider the nearest and next nearest neighbor (NNN) interaction with strengths $V_{1}$ and $V_{2}$, respectively. In the present study, we consider a three-body hardcore constraint $(\hat{a}^{\dagger}_i)^3=0$, which can occur for dominant three-body interactions and loss processes \cite{PSF_square_lattice_QMC_Wessel,PSF_square_lattice_QMC_K_Ng,
Lewenstein_PRA,Yang_PRA,Daley_PRL}. As a consequence, the boson occupancy at each lattice site is restricted upto $n_{\rm{max}}=2$. Moreover, such three-body constraint favors the pairing of bosons.

\subsection{Mean field Gutzwiller approach}
\label{MF_theory}
First, we revisit the system under the single-site mean field (MF) approximation to understand the essential qualitative features of the present model and the possible phases. In absence of the inter-site correlations, the single-site MF Hamiltonian can be written as,
\begin{eqnarray}
\hat{\mathcal{H}}^{i}_{\rm MF} &=& -\mu\hat{n}_{i} + \frac{U}{2}\hat{n}_{i}(\hat{n}_{i}-1)+V_{1}\hat{n}_{i}m_{i} + V_{2}\hat{n}_{i}m'_{i}\notag \\
&&-t(\hat{a}^\dagger_{i}\phi^{a}_{i}+\hat{a}_{i}\phi^{a*}_{i}) -t_p(\hat{a}^{\dagger2}_{i}\phi^{p}_{i}+\hat{a}^{2}_{i}\phi^{p*}_{i})
\label{MF_Hamiltonian}
\end{eqnarray}
where $\phi^{a}_{i} = \sum^{\rm NN}_{j}\langle \hat{a}_{j} \rangle$ and $\phi^{p}_{i} = \sum^{\rm NN}_{j}\langle \hat{a}^2_{j} \rangle$ denote the atomic superfluid (ASF) and pair superfluid (PSF) order parameters, respectively. On the other hand, the MF density at NN and NNN sites of the $i^{\rm{th}}$ site are denoted by $m_{i} = \sum^{\rm NN}_{j} \langle \hat{n}_{j} \rangle$ and ${m}'_{i} = \sum^{\rm NNN}_{j'} \langle \hat{n}_{j'} \rangle$, respectively. Equivalently, at zero temperature, the ground state can also be described by the Gutzwiller variational wavefunction $\ket{\Psi} = \prod_{i}\ket{\psi_{i}}$ \cite{P_Zoller_original,Krauth}, where the wavefunction $\ket{\psi_{i}}$ at the $i^{\mathrm{th}}$ site can be written as,
\begin{eqnarray}
\ket{\psi_{i}} = \sum_{n}f^{(n)}_{i}\ket{n_{i}}
\label{gutzwiller_wavefunction}
\end{eqnarray}
Here, $f^{(n)}_{i}$ are complex variational amplitudes, satisfying the constraint $\sum_{n} |f^{(n)}_{i}|^2 = 1$. Moreover, in presence of the three-body hardcore constraint $(\hat{a}^\dagger_{i})^3=0$, the local basis vector $\ket{n_{i}}$ is restricted upto $n_{i}=2$. The dynamics of the system can be captured from the equations of motion (EOM) of the time dependent Gutzwiller amplitudes $f^{(n)}_{i}(t)$, which are given by,
\begin{align}
&\imath \dot{f}^{(n)}_{i} \!=f^{(n)}_{i}\left[\frac{U}{2}n\left(n-1\right)-\left(\mu-V_{1}m_{i}-V_{2}m'_{i}\right)n-\lambda_{i}\right] \notag\\
&\!-t_p\left(\phi^{p*}_{i}\sqrt{(n+2)(n+1)}f^{(n+2)}_{i}+\phi^{p}_{i}\sqrt{n(n-1)}f^{(n-2)}_{i}\right) \notag\\
&\!-t\left(\phi^{a*}_{i}\sqrt{(n+1)}f^{(n+1)}_{i}+\phi^{a}_{i}\sqrt{n}f^{(n-1)}_{i}\right)
\label{EOM_bosons}
\end{align}
where the different order parameters can be written in terms of the  variational amplitudes, 
\begin{subequations}
\begin{eqnarray}
\phi^{a}_{i}&=&\sum_{j,n}\!\sqrt{(n+1)}f^{(n+1)}_{j}\!f^{*(n)}_{j}\\
\phi^{p}_{i}&=&\sum_{j,n}\!\sqrt{(n+2)(n+1)}f^{(n+2)}_{j}\!f^{*(n)}_{j}\\
m_{i}&=&\sum_{j,n}n|f^{(n)}_{j}|^2; \quad
m'_{i}=\sum_{j',n}n|f^{(n)}_{j'}|^2
\end{eqnarray}
\end{subequations}
with $j(j')$ being the NN (NNN) sites of the $i^{\rm th}$ site. 
Due to the normalization condition $\sum_{n} |f^{(n)}_{i}|^2 = 1$, we introduce the Lagrange multiplier $\lambda_{i}$ at each lattice site, which can be obtained from the steady states $\bar{f}^{(n)}_{i}$ of Eq.~\eqref{EOM_bosons}, describing the ground state.
In order to obtain the excitation spectrum, the time evolution of small fluctuations $\delta f^{(n)}_{i}(t)$ around the steady states $\bar{f}^{(n)}_{i}$ is considered by linearizing the Eq.~\eqref{EOM_bosons} \cite{Sinha_EPL,Sinha_unpublished}.
By decomposing $\delta f^{(n)}_{i}(t)$ into Fourier modes, $\delta f^{(n)}_{i}(t)=e^{\imath\big(\vec{k}.\vec{r}_{i}-\omega(\vec{k}) t\big)}\delta f^{(n)}(\vec{k})$, we obtain the set of fluctuation equations in momentum space, which in turn yields the excitation spectrum $\omega(\vec{k})$ for various equilibrium phases. Apart from such variational approach, excitations can also be obtained by using  more involved numerical techniques, which include `Quantum Monte Carlo' (QMC) \cite{exc_spectra_QMC1,exc_spectra_QMC2,exc_spectra_QMC3}, Green's function method \cite{exc_spectra_Pelster}, as well as `Time Evolution Block Decimation' (TEBD) \cite{exc_spectra_TEBD} and time dependent matrix product states (MPS) \cite{exc_spectra_tMPS} etc. Unlike the above mentioned computationally complex techniques, which yield more accurate results, the time dependent Gutzwiller approach neglects the details of correlation, however, its computational simplicity is the main advantage which also allows us to obtain the excitations analytically, providing more physical insight. Moreover, this Gutzwiller approach has recently been extended to incorporate the higher order quantum fluctuations, which can be used to obtain more accurate results \cite{quantum_gutzwiller_Carusotov}.

Similar to the Gutzwiller wavefunction at zero temperature, within the mean field approximation, the density matrix at finite temperatures $T=1/\beta$ can be written in product form, $\hat{\rho}_{T}=\prod_{i} \hat{\rho}_{i}$, where $\hat{\rho}_{i}=e^{-\beta\hat{\mathcal{H}}^{\rm MF}_{i}}/Z_{i}$ describes the density matrix at the $i^{\rm th}$ site and the corresponding partition function is $Z_{i}={\rm Tr}(e^{-\beta\hat{\mathcal{H}}^{\rm MF}_{i}})$ for the MF Hamiltonian $\hat{\mathcal{H}}^{\rm MF}_{i}$ given in Eq.~\eqref{MF_Hamiltonian}. The thermal average of any local observable $\hat{\mathcal{O}}_{i}$ can be obtained from the relation, $\langle \mathcal{\hat{O}}_{i} \rangle = {\rm Tr}(\hat{\mathcal{O}}_{i}\hat{\rho}_{i})$, and the order parameters introduced in the mean field Hamiltonian can be obtained self-consistently at finite temperatures using the above description.

Under single-site MF approximation, the inter-site correlations are neglected, which can be implemented in a systematic manner using the cluster mean field (CMF) approach, as described in the next section.

\subsection{Cluster mean field theory}
\label{CMF_theory}
\begin{figure}
	\centering
	\includegraphics[width=0.8\columnwidth]{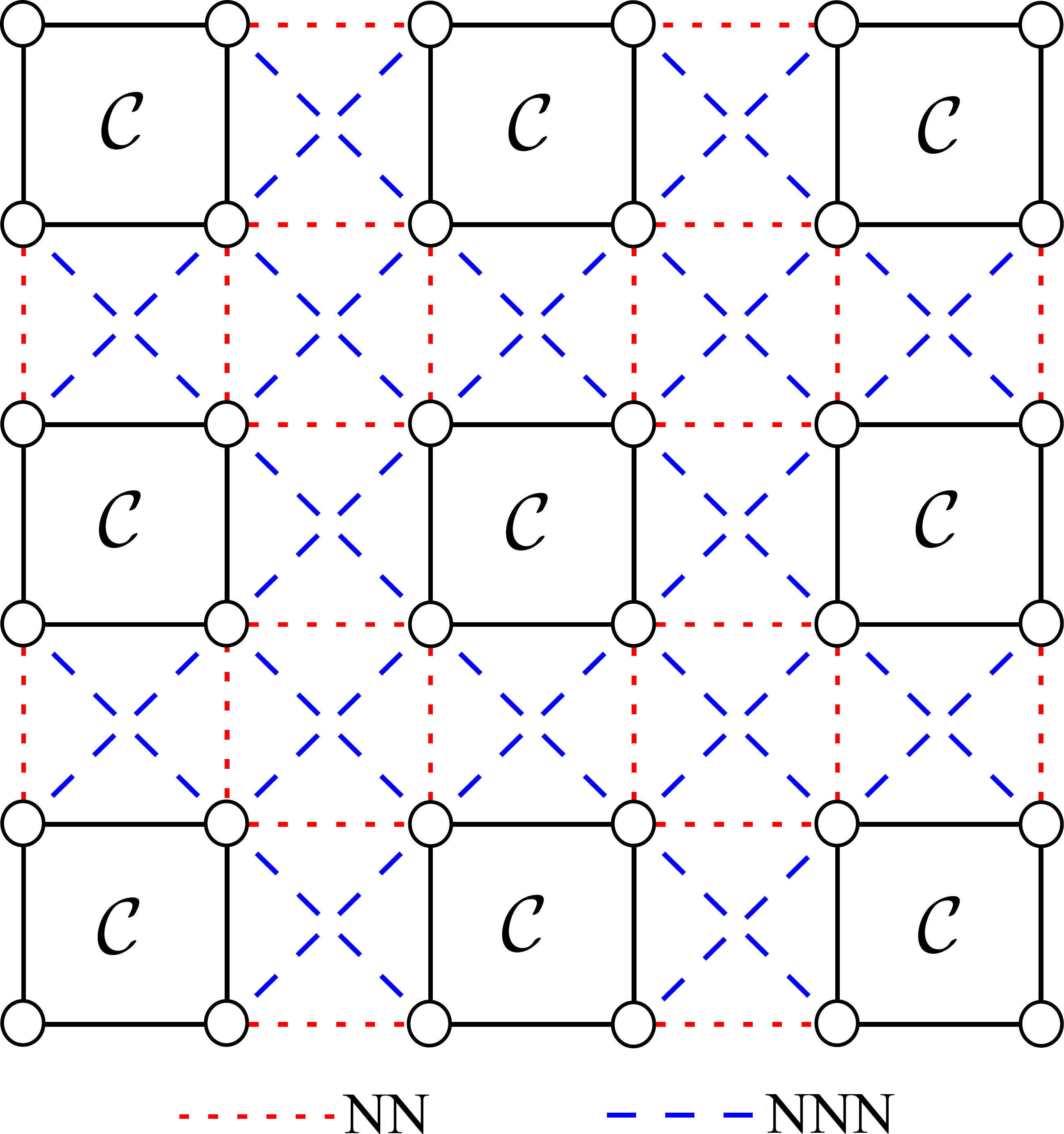}
	\caption{Schematic of $2\times 2$ clusters $\mathcal{C}$: The boundaries of each cluster $\mathcal{C}$ are denoted by solid black lines, within which the correlations are considered exactly.  The dotted red (dashed blue) lines correspond to the NN (NNN) interactions between the clusters, which are treated at the mean field level, representing the mean field quantities $\phi^{a}_{i}$, $\phi^{p}_{i}$, and $m_{i}$ ($m'_{i}$), as described by the Hamiltonian $\mathcal{H}^{i}_{\rm MF}$ in Eq.~\eqref{mf_ham}.}      
\label{Fig1}
\end{figure}

In this section, we employ the cluster mean field (CMF) method \cite{Yamamoto_triangular_CMFT, Yamamoto_square_CMFT,Heydarinasab_CMFT,Demler_CMFT,Luhmann_CMFT,
Fazio_CMFT,Suthar_CMFT_1,Suthar_CMFT_2,Suthar_CMFT_3,Manali_CMFT_1,
Sayak_CMFT,Manali_CMFT_2}, under which the inter-site correlations are taken into account to capture the underlying features of the present system beyond the single-site mean field (MF) approach. Unlike the single-site MF, in the CMF technique, the inter-site correlations between the bosons within a given cluster $\mathcal{C}$ are treated exactly, by using the method of exact diagonalization. The Hamiltonian corresponding to the cluster $\mathcal{C}$ can be written in two parts, 
\begin{eqnarray}
\hat{\mathcal{H}} = \hat{\mathcal{H}}_{\mathcal{C}} + \hat{\mathcal{H}}_{\rm{MF}}
\label{CMF hamiltonian_1}
\end{eqnarray}
where $\hat{\mathcal{H}}_{\mathcal{C}}$ describes the exact Hamiltonian of the interacting bosons within the cluster $\mathcal{C}$, and the Hamiltonian $\hat{\mathcal{H}}_{\rm{MF}}$ describes the MF interactions for the sites at the edge of the cluster $\mathcal{C}$, 
\begin{eqnarray}
\hat{\mathcal{H}}_{\rm{MF}}&=&\!\!\!\!\sum_{i\, \in\, {\rm edge\, sites}}\!\!\!\!\hat{\mathcal{H}}^i_{\rm{MF}} \\
\label{CMF hamiltonian_2}
\hat{\mathcal{H}}^{i}_{\rm MF}&=&-t(\hat{a}^\dagger_{i}\phi^{a}_{i}+\hat{a}_{i}\phi^{a*}_{i})-t_p(\hat{a}^{\dagger 2}_{i}\phi^{p}_{i}+\hat{a}^{2}_{i}\phi^{p*}_{i}) \notag\\
 &&+ V_{1}\hat{n}_{i}m_{i} + V_{2}\hat{n}_{i}m'_{i} 
\label{mf_ham}
\end{eqnarray}
The MF order parameters are given by, 
\begin{subequations} 
\begin{align}
&\phi^{a}_{i}=\sum^{\rm NN}_{j \notin \mathcal{C}} \langle \hat{a}_{j}\rangle,\quad \phi^{p}_{i}=\sum^{\rm NN}_{j \notin \mathcal{C}} 
\langle \hat{a}^2_{j}\rangle \\ 
&m_{i}=\sum^{\rm NN}_{j \notin \mathcal{C}} \langle \hat{n}_{j}\rangle, \quad  m'_{i}=\sum^{\rm NNN}_{j' \notin \mathcal{C}} \langle \hat{n}_{j'}\rangle
\end{align}
\label{schematic_equation}
\end{subequations} 
where $j$ ($j'$) denotes the NN (NNN) lattice sites outside the cluster $\mathcal{C}$. In Fig.\ref{Fig1}, we show the schematic of $2\times 2$ clusters $\mathcal{C}$. The solid black lines denote the boundaries of each cluster $\mathcal{C}$, within which the correlations are considered exactly by using the Hamiltonian $\hat{\mathcal{H}}_{\mathcal{C}}$ in Eq.~\eqref{CMF hamiltonian_1}. Whereas, the dotted red (dashed blue) lines correspond to the NN (NNN) interactions between the clusters, which are treated at the mean field level, representing the mean field quantities $\phi^{a}_{i}$, $\phi^{p}_{i}$, and $m_{i}$ ($m'_{i}$), as described by the Hamiltonian $\mathcal{H}^{i}_{\rm MF}$, given in Eq.~\eqref{mf_ham}. Note that, due to the presence of sub-lattice symmetry throughout the entire lattice plane, all the clusters can be considered equivalent. Consequently, the MF quantities $\langle \hat{\mathcal{O}} \rangle$ can be obtained from the lattice sites of the given cluster $\mathcal{C}$.  At zero temperature, the average value of any observable $\hat{\mathcal{O}}$ can be obtained from  $\langle \hat{\mathcal{O}} \rangle=\langle\psi|\hat{\mathcal{O}}|\psi\rangle$, where $|\psi\rangle$ represents the ground state wavefunction of the Hamiltonian $\hat{\mathcal{H}}$ in Eq.~\eqref{CMF hamiltonian_1} corresponding to cluster $\mathcal{C}$. The order parameters both at zero and finite temperatures can be obtained self-consistently using the prescription of MF theory.
For phases at finite temperature, the thermodynamic quantities can be computed from, 
\begin{eqnarray}
\langle \mathcal{\hat{O}} \rangle = {\rm Tr}(\hat{\mathcal{O}}\hat{\rho}_{T})
\label{thermal_average_MF}
\end{eqnarray} 
using the equilibrium density matrix  $\hat{\rho}_{T}=e^{-\beta\hat{\mathcal{H}}}/Z$ at temperature $T=1/\beta$ obtained in a self-consistent manner, where $Z$ denotes the total partition function of the cluster $\mathcal{C}$.

\begin{figure*}
	\centering
	\includegraphics[width=\textwidth]{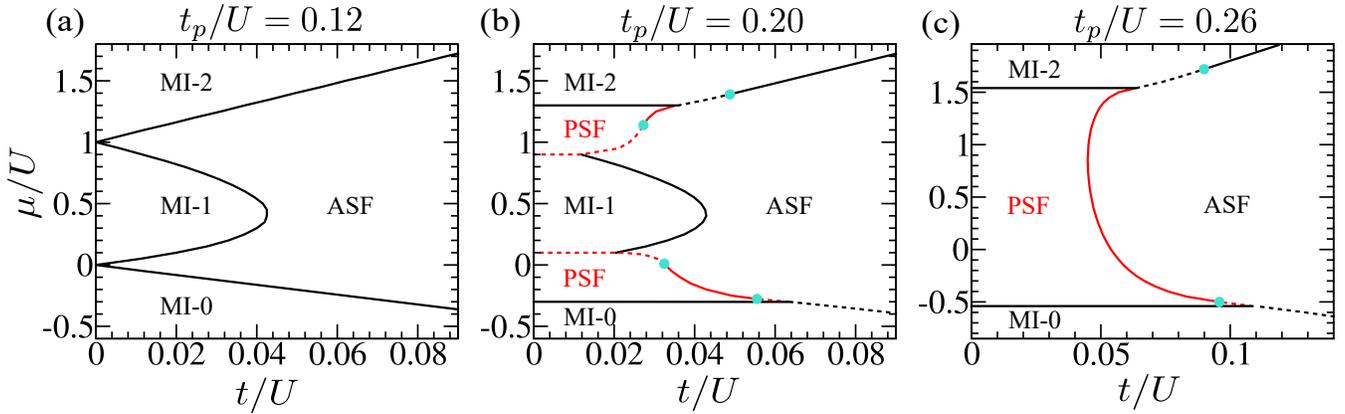}
	\caption{Mean field (MF) phase diagrams in the $\mu/U$-$t/U$ plane for increasing values of pair hopping strength $t_{p}/U$. The solid lines correspond to the continuous transitions, whereas the dotted lines represent the first-order transitions. The filled circular symbols denote the tri-critical points, where the first-order and continuous phase boundaries meet.}      
\label{Fig2}
\end{figure*}
 
Due to the inclusion of correlations in CMF theory, the quantitative accuracy of the transition points increases with increasing cluster size. In addition, the finite cluster-size scaling analysis can yield results which are in close agreement with that of the quantum Monte Carlo (QMC) simulations \cite{Yamamoto_triangular_CMFT,Yamamoto_square_CMFT,Manali_CMFT_1,
Sayak_CMFT}. Although the phase boundaries can be obtained quite accurately, the critical behavior of the transition may not be captured from CMF technique, particularly for BKT transition \cite{BKT_transition}, which occurs for finite temperature melting of superfluid to normal fluid in 2D.

In the subsequent sections, we study the different equilibrium phases of the present model and obtain both the zero as well as finite temperature phase diagrams by using the MF and CMF approach.

\section{Formation of pair superfluid phase}
\label{Formation_of_pair_SF_phase}
In this section, we investigate the possible phases of bosons on a square lattice in presence of the pair hopping, particularly focusing on the formation of the pair superfluid (PSF) phase. For this purpose, we consider the following Hamiltonian without any long-range interactions,
\begin{eqnarray}
\hat{\mathcal{H}} = \hat{\mathcal{H}}_{\rm BHM} + \hat{\mathcal{H}}_{\rm p}
\label{Ham_pairing}
\end{eqnarray}
where $\hat{\mathcal{H}}_{\rm BHM}$ and $\hat{\mathcal{H}}_{\rm p}$ are described previously in Sec.\ref{Model_and_Method}. Apart from the various Mott insulating phases, two different kinds of superfluid, namely the atomic superfluid (ASF) and the pair superfluid (PSF) can appear in the phase diagram as a result of the interplay between single particle hopping $t$ and pair hopping $t_{p}$. The ASF phase is characterized by $\phi^{a}=\langle\hat{a}\rangle \neq 0$, whereas the PSF phase can be identified with $\phi^{p}=\langle\hat{a}^2\rangle \neq 0$, but $\phi^{a} = 0$. Note that, a non-vanishing PSF order $\phi^{p} \neq 0$ can exist in the ASF phase. As a consequence of the pair hopping term, two bosons can simultaneously move to their nearest neighbor sites, resulting in a pair superfluidity. Whereas, the single particle hopping can break such pairs. In the following, we describe the features of the zero temperature phase diagrams obtained by tuning the pair hopping strength, using both the MF and CMF approach.

\begin{figure}[b]
	\centering
	\includegraphics[width=1.02\columnwidth]{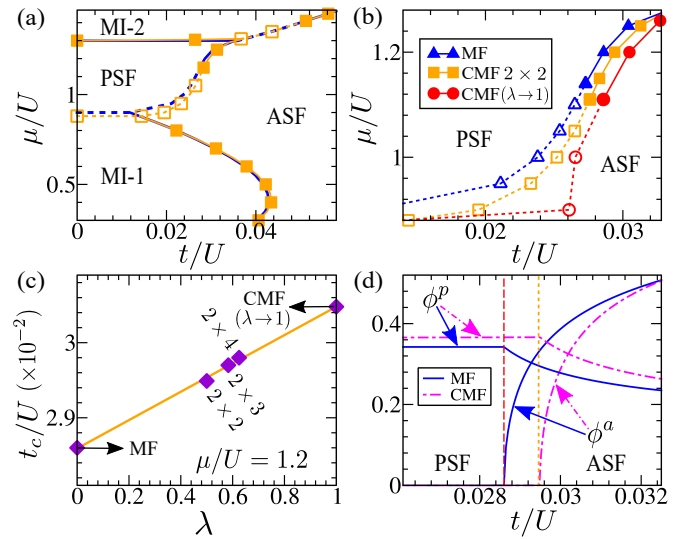}
	\caption{(a) Comparison between mean field (MF) and cluster mean field (CMF) phase diagram around the particle dominated region in the $\mu/U$-$t/U$ plane for pair hopping strength $t_{p}/U=0.20$. The thick solid (dotted) lines correspond to the continuous (first-order) transition within MF. The filled (open) squares denote the continuous (first-order) boundaries obtained via $2 \times 2$ cluster, where the thin solid (dotted) lines act as a guide to the eye. (b) Phase boundaries zoomed near the PSF-ASF transition above the MI-1 lobe, where triangles, squares, and circles are obtained from the MF, CMF-$2 \times 2$, and cluster-size extrapolation at $\lambda\!\!\rightarrow\!\!1$ \cite{cluster_note}, respectively. (c) Finite cluster-size scaling to extrapolate the critical point of the continuous PSF-ASF transition at $\lambda\!\!\rightarrow\!\!1$. (d) Variation of ASF and PSF order  parameters $\phi^{a}$ and $\phi^{p}$ at $\mu/U =1.2$, respectively, with increasing single particle hopping strength $t$ using the MF and CMF ($2\times2$) method. The vertical dashed (dotted) line denotes the continuous PSF-ASF transition point obtained from MF (CMF) theory.}      
\label{Fig3}
\end{figure}
  
To this end, we elaborate the different features of the zero temperature phase diagrams in the $\mu/U$-$t/U$ plane, for different regimes of pair hopping amplitude $t_p$. The regime with sufficiently small $t_{p}$ is considered first, for which the phase diagram corresponds to the usual BHM. However, due to the three-body constraint, we obtain the Mott insulating phases upto filling $n_{0}=2$, surrounded by the ASF phase, as shown in Fig.\ref{Fig2}(a).
Above a certain critical value of pair hopping strength $t^{c1}_{p} \simeq U/2z$ (with $z$ being the coordination number), the PSF phase emerges between the Mott phases for sufficiently small single particle hopping strength $t$, as depicted in Fig.\ref{Fig2}(b). The PSF-ASF phase boundary exhibits intriguing behavior, where the first and second-order lines meet together at the `tri-critical' points. Such PSF-ASF transition will be discussed later in details in Sec.\ref{PSF_ASF_transition}. By further increasing $t_{p}$, the Mott-1 (MI-1) phase disappears above a critical value $t^{c2}_p \simeq U/z$, and is replaced by the PSF phase, formed in a large region between MI-0 and MI-2 phases [see Fig.\ref{Fig2}(c)].  As seen from Fig.\ref{Fig2}(b,c), the boundaries between PSF and insulating MI-0 as well as MI-2 phase are independent of $t$, which is an artifact of single-site MF theory.

Although the qualitative features of the phase diagram can be obtained from the simple MF theory, to incorporate the effect of correlations systematically, we also study the formation of PSF phase using the CMF theory. We obtain the phase diagram in the particle dominated region within the CMF method focusing on the formation of PSF between MI-1 and MI-2 phase, which is compared with the MF result, as depicted in Fig.\ref{Fig3}(a). Now, the boundary between PSF and MI-2 phase exhibits weak dependency on the single particle hopping strength $t$. In Fig.\ref{Fig3}(b), we show how the PSF-ASF phase boundary shifts by inclusion of larger cluster sizes. The shift in the critical point can be calculated more accurately by using finite cluster-size scaling \cite{Yamamoto_triangular_CMFT,Yamamoto_square_CMFT}, as shown in Fig.\ref{Fig3}(c). The variation of the order parameters $\phi^{a}$ and $\phi^{p}$ across the PSF-ASF transition is depicted in Fig.\ref{Fig3}(d). Note that, the PSF order parameter $\phi^{p}$ remains non-vanishing in the ASF phase, however its behavior changes across the transition. 

We have also studied the PSF-ASF transition and obtained the phase diagrams by varying the pair hopping strength $t_{p}$, which we discuss in Appendix \ref{app1}.

\subsection{Collective excitations}
\label{collective_excitations_PSF}
In this subsection, we study the low-energy collective modes of different phases at zero temperature using the method of linear stability analysis, as outlined in Sec.\ref{MF_theory}. The excitation spectrum not only captures the characteristic features of the different phases, but it is also useful for detecting the transitions between them.

We begin our analysis starting from the atomic limit ($t \sim 0$), to understand how the inclusion of pair hopping term $\hat{\mathcal{H}}_{\rm p}$ modifies the phase diagram of the usual BHM. The single-particle and hole excitation of the MI-0 and MI-2 phases are given by $\epsilon^{1p}=-\mu$ and $\epsilon^{1h}=\mu-U$, respectively, which vanish at $\mu^{\rm L}_{c}$(MI-1) $=0$ and $\mu^{\rm U}_{c}$(MI-1) $=U$, giving rise to the MI-1 phase. Although the pair particle and hole excitations cost higher energy, the inclusion of pair hopping term leads to the de-localization of such excitations, resulting in a reduction of their energies. Using the degenerate perturbation theory, the pair particle and hole excitation of the MI-0 and MI-2 phases become $\epsilon^{2p} = U-2\mu-2zt_{p}$ and $\epsilon^{2h} = 2\mu-U-2zt_{p}$, respectively. Above a critical value of pair hopping strength $t^{c1}_{p}=U/2z$, such pair excitations become unstable at $\mu^{\rm L}_{c}({\rm PSF})=U/2-zt_{p}$ and $\mu^{\rm U}_{c}({\rm PSF})=U/2+zt_{p}$ before the instability of single particle and hole excitations i.e. when $\mu^{\rm L}_{c}({\rm PSF})<\mu^{\rm L}_{c}$(MI-1) and $\mu^{\rm U}_{c}({\rm PSF})>\mu^{\rm U}_{c}$(MI-1). The instability of such de-localized pair excitations lead to the formation of the PSF phase for $t_{p} \gtrsim t^{c1}_{p}$.

For finite hopping strength $t$, we obtain the dispersion relation $\omega(\vec{k})$ of the excitations for both the insulating and the superfluid phases by linearizing the Eq.~\eqref{EOM_bosons} [describing the equations of motion of the Gutzwiller amplitudes], as outlined in Sec.\ref{MF_theory}. Following the linear stability analysis, we compute the excitation branches for the MI-0 phase at finite $t$, given by,
\begin{subequations}
\begin{align}
\omega^{s}(\vec{k}) &= 2t\epsilon(\vec{k})+\mu \label{MI-0_s_branch}\\
\omega^{p}(\vec{k}) &= 4t_{p}\epsilon(\vec{k})-U+2\mu \label{MI-0_p_branch}
\end{align}
\label{MI-0_excitations}
\end{subequations}
where $\epsilon(\vec{k}) = \sum^{d}_{l=1}\cos{k_{l}}$ and $d=z/2$ is the dimension of the lattice. Similarly, the excitation spectrum for the MI-2 phase at finite $t$ is given by,
\begin{subequations}
\begin{align}
\omega^{s}(\vec{k}) &= 4t\epsilon(\vec{k})-\mu+U \label{MI-2_s_branch}\\
\omega^{p}(\vec{k}) &= 4t_{p}\epsilon(\vec{k})+U-2\mu \label{MI-2_p_branch}
\end{align}
\label{MI-2_excitations}
\end{subequations} 
Note that, in the atomic limit ($t \sim 0$), the excitation branches $\omega^{s/p}$ reduce to the single and pair particle/hole excitations, respectively.
The instability of the pair excitations gives rise to the phase boundaries for the PSF to MI-0/MI-2 transition, which are independent of $t$ at the level of MF. On the other hand, the vanishing of the gap of $\omega^{s}$ branch in Eq.~\eqref{MI-2_s_branch} gives the continuous transition line between ASF to MI-2 phase [see Fig.\ref{Fig2}(b,c)]. The gapped excitations of the MI-1 phase are independent of $t_{p}$ and are given by,
\begin{align}
\omega^{\pm}(\vec{k}) &=  U/2-\mu-t\epsilon(\vec{k})\!\pm\!\sqrt{t^2\epsilon^2(\vec{k})+U^2/4-3Ut\epsilon(\vec{k})}
\end{align}
From the condition $\omega^{\pm}=0$, the phase boundary of the  continuous transition between MI-1 to ASF phase is obtained [see the thick black line between MI-1 and ASF in Fig.\ref{Fig2}(a,b)].

\begin{figure}
	\centering
	\includegraphics[width=1.02\columnwidth]{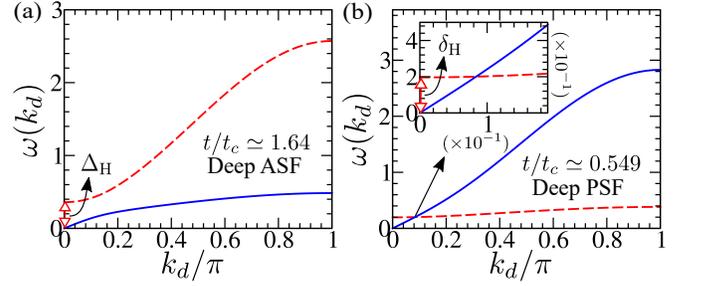}
	\caption{ASF-PSF transition with decreasing single particle hopping $t$ for fixed chemical potential $\mu/U=-0.1$ and pair hopping $t_{p}/U = 0.20$: low-energy collective excitations $\omega(k_{d})$ of the deep (a) ASF and (b) PSF phase along the diagonal of the Brillouin zone $k_{x}=k_{y}=k_{d}$. The inset in (b) illustrates the low-energy excitations zoomed near $k_{d}\sim0$, indicating level crossing between the modes. The critical point for PSF to ASF transition is $t_{c}/U\simeq 0.0364$.}      
\label{Fig4}
\end{figure}

Next, we focus on the excitations of the superfluid phases. In the ASF phase, the usual gapless sound mode exists due to the $U$(1) symmetry breaking, however unlike the case of hardcore bosons, we obtain the gapped Higgs mode of the ASF phase \cite{Higgs_Altman,Higgs_review,Higgs_expt_SF_1,Higgs_expt_SF_2}, even in the presence of the three-body constraint, as shown in Fig.\ref{Fig4}(a). 
The energy gap of the Higgs mode $\Delta_{\rm H}$ can be evaluated at $\vec{k}=0$ of its dispersion relation, as illustrated in Fig.\ref{Fig4}(a). Such gap of the Higgs mode in the usual Bose-Hubbard model vanishes at the tip of the Mott lobe corresponding to the particle hole symmetry.
Close to the ASF-PSF transition, where the single and pair particle hopping become comparable, an avoided level crossing can occur between the two modes, indicating hybridization between them. 
As depicted in Fig.\ref{Fig4}(b), the PSF phase also has a gapless sound mode ($\omega^{\rm G}$) as well as a  gapped excitation ($\omega^{\rm H}$), which can be obtained analytically,
\begin{subequations}
\begin{align}
\omega^{\rm G}(\vec{k}) &= \sqrt{(\mathcal{A}^2-\mathcal{B}^2)+\frac{(\mathcal{C}^2+\mathcal{D}^2)}{2}+\sqrt{\mathcal{A}^2(\mathcal{C}+\mathcal{D})^2+\Gamma_{1}}}\\
\omega^{\rm H}(\vec{k}) &=\sqrt{\!\left(\!2(t\epsilon(\vec{k})\!-\!zt_{p})\cos^2\!{\bar{\theta}}\!+\!4t\epsilon(\vec{k})\sin^2\!{\bar{\theta}}\!+\!U\!-\!\mu\right)^2\!\!-\!\Gamma_{2}} 
\end{align}
\label{psf_excitations}
\end{subequations}
where $\Gamma_{1}=(\mathcal{C}-\mathcal{D})^2[(\mathcal{C}+\mathcal{D})^2-4\mathcal{B}^2]/4$, $\Gamma_{2}=8t^2\epsilon^2(\vec{k})\sin^2\!{2\bar{\theta}}$, $\mathcal{A}\!=\!zt_{p}\sin2\bar{\theta}$, $\mathcal{B}\!=\!2t_{p}\epsilon(\vec{k})\sin2{\bar{\theta}}$, $\mathcal{C}\!=\!4t_{p}\epsilon(\vec{k})\sin^2\!{\bar{\theta}}+U/2-\mu-zt_{p}, \mathcal{D}\!=\!2t_{p}(2\mathcal{\epsilon}(\vec{k})-z)\cos^2\!{\bar{\theta}}$, and $\cos{2\bar{\theta}}\!=\!(U-2\mu)/2zt_{p}$. Note that, the sound velocity of the PSF phase becomes $v_{p} = \sqrt{2}zt_{p}\phi_{p}$, which scales with $t_{p}$ and PSF order $\phi_{p}$.
Unlike the ASF phase, a level crossing occurs between the two modes in the PSF phase.
In the PSF phase, the gap of the higher energy excitation $\delta_{\rm H}=\omega^{\rm H}(\vec{k}=0)$ [see Fig.\ref{Fig4}(b)] provides the stability of the bosonic pairs, which decreases with increasing the single particle hopping $t$, since it leads to the pair breaking.

\subsection{PSF-ASF transition}
\label{PSF_ASF_transition}
\begin{figure}
	\centering
	\includegraphics[width=1.02\columnwidth]{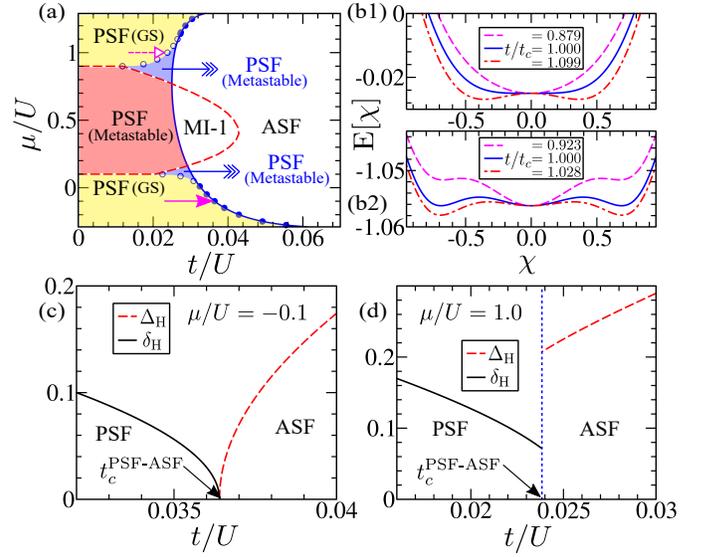}
	\caption{(a) MF phase diagram in the $\mu/U$-$t/U$ plane for $t_{p}/U=0.20$, illustrating the different regions of stable PSF phase either as ground or metastable state. The PSF phase coexists as a metastable state with MI-1 and ASF phases in the red and blue shaded  regions, respectively. In contrast, PSF exists as a ground state (GS) in the yellow shaded regions. Variation of the Landau-Ginzburg (LG) energy E$[\chi]$ across (b1) continuous PSF to ASF transition at $\mu/U=-0.1$ [shown by pink solid arrow in (a)] and (b2) first-order transition at $\mu/U=1.0$ [shown by pink dashed white headed arrow in (a)]. Variation of the energy gap of the gapped excitation, $\Delta_{\rm H}$ (for ASF) and $\delta_{\rm H}$ (for PSF) across the (c) continuous and (d) first-order PSF to ASF transition. The critical point of the continuous (first-order) PSF-ASF transition is $t_{c}/U\simeq 0.0364(0.0238)$.}      
\label{Fig5}
\end{figure}
\begin{figure*}
	\centering
	\includegraphics[width=\textwidth]{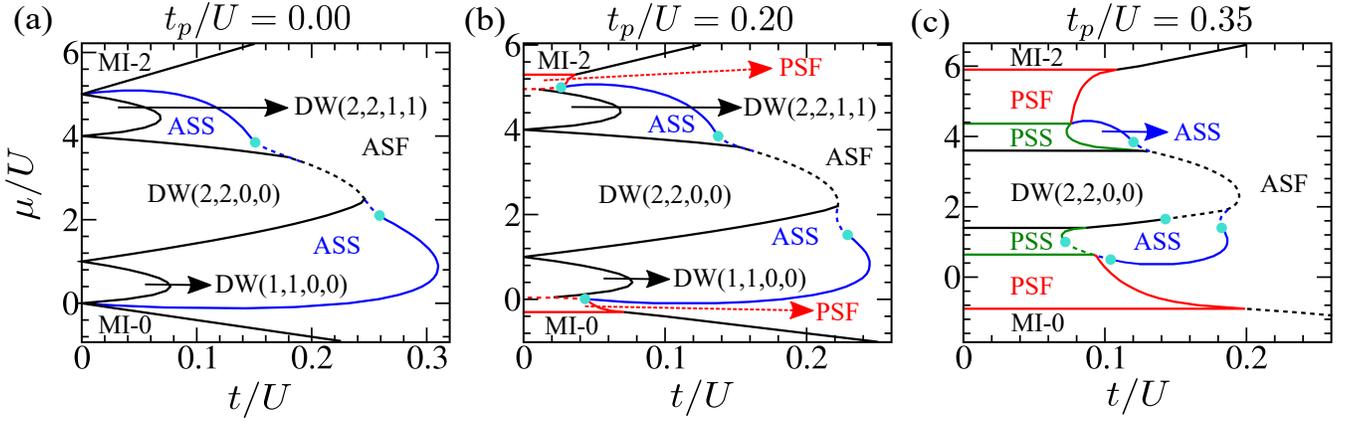}
	\caption{MF phase diagrams in the $\mu/U$-$t/U$ plane with increasing values of pair hopping strength $t_{p}/U$ for next nearest neighbor (NNN) interaction strength $V_{2}/U=0.5$. The solid lines correspond to the continuous transitions, whereas the dotted lines represent the first-order transitions. The filled circular symbols denote the tri-critical points, where the nature of the transition changes from first-order to continuous. The density structure is represented by ($n_{1}$,$n_{2}$,$n_{3}$,$n_{4}$) where $n_{i}$ denotes the site occupancies on a unit cell in the anti-clockwise direction.}      
\label{Fig6}
\end{figure*}
To this end, we analyze the PSF-ASF transition in more details, which reveals intriguing behavior. To understand the nature of this transition from energetics consideration, we compute the energy density E using simple variational wavefunction, 
\begin{eqnarray}
\ket{\psi}_{i} = \cos{\chi}(\cos{\theta\ket{0}_{i}}+\sin{\theta}\ket{2}_{i})+\sin{\chi}\ket{1}_{i} 
\end{eqnarray}
which describes the PSF phase for $\bar{\chi}=0$ and $\cos{2\bar{\theta}}=(U-2\mu)/2zt_{p}$. The corresponding energy density E is given by,
\begin{align}
{\rm E}[\chi,\theta] =& -\frac{zt}{2}\sin^2\!{2\chi}\left(\sin{\theta}+\frac{\cos{\theta}}{\sqrt{2}}\right)^2\!\!-\frac{zt_{p}}{2}\cos^4\!{\chi}\sin^2\!{2\theta}\notag\\
&+(U-2\mu)\cos^2\!{\chi}\,\sin^2\!{\theta}-\mu\sin^2\!{\chi}
\label{free_energy_density}
\end{align}
where different phases can be obtained by minimizing E. From the condition of local minima of E, we obtain the region in parameter space, where PSF can exist as a stable phase corresponding to either metastable or ground state. The boundary line [see the blue solid line in Fig.\ref{Fig5}(a)] embedding the region of stable PSF phase is given by,
\begin{align}
&-2(zt+U-2\mu)\sin^2\!{\bar{\theta}}-2(zt+\mu)+2zt_{p}\sin^2\!{2\bar{\theta}}\notag\\
&-2\sqrt{2}zt\sin{2\bar{\theta}}=0
\label{PSF_stable_boundary}
\end{align}
where $\cos{2\bar{\theta}}=(U-2\mu)/2zt_{p}$. The stable PSF phase can coexist with other phases (like MI-1 and ASF) as a metastable state, which is shown by the red and blue shaded region in Fig.\ref{Fig5}(a). Whereas, the PSF phase in the yellow shaded region of the phase diagram exists as a true ground state. Note that, for continuous transition, the above-mentioned boundary in Eq.~\eqref{PSF_stable_boundary} coincides with the PSF-ASF phase boundary obtained from global minima of the energy. We can understand this situation in a better way by constructing the Landau-Ginzburg (LG) energy E$[\chi]$, which is obtained by minimizing the energy density in Eq.~\eqref{free_energy_density} for fixed values of $\chi$. Note that, the present model has $U(1) \times \mathbb{Z}_{2}$ symmetry \cite{PSF_square_lattice_QMC_Wessel,PSS_triangular_lattice_Zhang,Christophe_Mora_PRL}, where both the superfluid phases correspond to the broken $U(1)$ symmetry. For continuous PSF to ASF transition, the energy density takes a double well like structure, as shown in Fig.\ref{Fig5}(b1), where the minima correspond to the ASF phase with broken $\mathbb{Z}_{2}$ symmetry. As a consequence, the gapped mode of both the ASF and PSF phases with energy gaps $\Delta_{\rm H}$ and $\delta_{\rm H}$, respectively [see Fig.\ref{Fig4}(a,b)], become gapless at the critical point, as shown in Fig.\ref{Fig5}(c). 
Such gap vanishing phenomena at the critical point for both the phases is a reflection of the continuous transition of Ising class related to $\mathbb{Z}_{2}$ symmetry breaking \cite{Subir_Sachdev}.
On the other hand, for the first-order transition, the LG energy takes the shape as depicted in Fig.\ref{Fig5}(b2). The higher energy local minima at $\chi=0$ describes the metastable PSF phase, which corresponds to the blue shaded region outside the MI-1 phase boundary in Fig.\ref{Fig5}(a). In this case, both the gaps of the higher energy excitation do not vanish at the critical point and exhibit a discontinuous jump, signifying the first-order nature of the transition, as evident from Fig.\ref{Fig5}(d). Here we point out that, for continuous transition, the phase boundaries of the PSF phase can also be obtained from the condition $\omega^{\rm H}=0$.

The present analysis of the low-energy collective modes provides a clear picture of the formation of the PSF phase as well as the nature of its transition to ASF, which can be useful for its identification. We have also studied the formation of the PSF phase as well as its excitation spectrum in the presence of attractive interaction, which is discussed in details in appendix \ref{app2}.

\section{Formation of Pair supersolid}
\label{Formation_of_pair_SS_phase}
In this section, we investigate the effect of long-range component $\hat{\mathcal{H}}_{\rm lr}$ of the interaction on the PSF phase, which can act as a key ingredient for density ordering. In this case, the system is described by the full Hamiltonian given in Eq.~\eqref{hamiltonian}. Since the bosonic pairs in the presence of three-body constraint act as effective hardcore bosons and stable supersolids of hardcore bosons can only form in the regime of dominant next nearest neighbor (NNN) interaction \cite{Yamamoto_square_CMFT,Batrouni_QMC_1,
Batrouni_QMC_2}, here for simplicity we only consider the effect of NNN interaction to investigate the formation of pair supersolid (PSS) phase. In the rest of the paper, we neglect the nearest neighbor (NN) interaction strength by setting $V_{1}=0$.

\begin{figure}[b]
	\centering
	\includegraphics[width=\columnwidth]{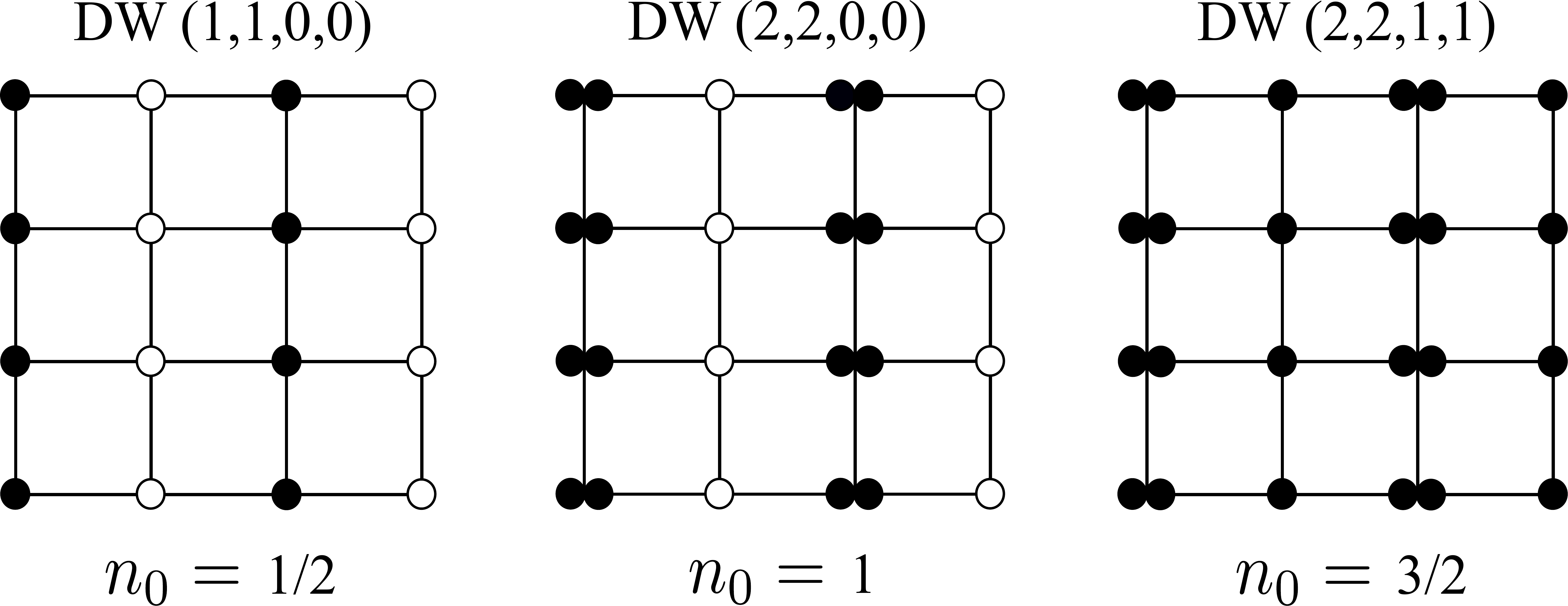}
	\caption{Schematic of different possible solids (density waves DW) with filling $n_{0}$ in the presence of only NNN interaction.  The density structure is represented by ($n_{1}$,$n_{2}$,$n_{3}$,$n_{4}$) where $n_{i}$ denotes the site occupancies on a unit cell in the anti-clockwise direction.}      
\label{Fig7}
\end{figure}

First, we analyze the system within the single-site MF theory to understand the possible density ordered phases. It is expected that the NNN interaction can give rise to stripe density ordering \cite{Yamamoto_square_CMFT,Batrouni_QMC_1,Batrouni_QMC_2} and we identify the regime of interaction where such stripe ordering can lead to the formation of PSS phase. The density structure can be represented by ($n_{1}$,$n_{2}$,$n_{3}$,$n_{4}$), where $n_{i}$ denotes the site occupancies on a unit cell in the anti-clockwise direction.
To gain the physical insight, we analyze the system ignoring both the single and pair hopping terms. At unit filling, the MI-1 phase is favoured for sufficiently small $V_{2}$ with energy density $E = -\mu+2V_{2}$. As $V_{2}$ increases, the MI-1 phase is replaced by unit filled density wave DW(2,2,0,0) with energy density $E = -\mu+U/2$, where bosonic pairs are aligned along a line, giving rise to stripe ordering. This can be viewed as a stripe pair solid which can melt to form a PSS phase due to the effect of the pair hopping. It is evident from the energetic consideration that, this pair density wave (DW) phase forms when $V_{2} > U/4$, therefore we only consider this regime in the rest of the work to study the formation of PSS phase.

\begin{figure}
	\centering
	\includegraphics[width=0.85\columnwidth]{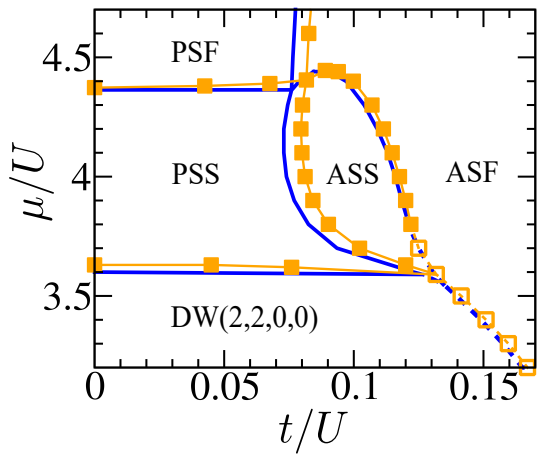}
	\caption{Cluster mean field Phase diagram in the $\mu/U$-$t/U$ plane for $t_{p}/U=0.35$ and $V_{2}/U=0.5$, obtained using $2 \times 2$ cluster. The thick solid lines correspond to the second-order transitions, whereas the thick dotted lines represent the first-order transitions within MF. The open (filled) squares denote the first (second)-order boundaries obtained via CMF theory where the thin dotted (solid) lines act as a guide to the eye. }      
\label{Fig8}
\end{figure}

We study the evolution of the phase diagram by increasing the pair hopping strength $t_{p}$. For vanishingly small $t_{p}$, in addition to DW(2,2,0,0), the solids DW(1,1,0,0) and DW(2,2,1,1) also appear with filling $1/2$ and $3/2$, respectively, which are surrounded by the atomic supersolid (ASS) phase [See Fig.\ref{Fig6}(a)]. A schematic of the different DW phases mentioned above is illustrated in Fig.\ref{Fig7}. For sufficiently strong single particle hopping $t$, the density ordering vanishes and a homogeneous atomic superfluid (ASF) phase is formed. With increasing $t_{p}$, the MI-0 and MI-2 phases melt to form the PSF phase [see Fig.\ref{Fig6}(b)], as discussed in the previous section Sec.\ref{Formation_of_pair_SF_phase}. Further increasing the pair hopping above a critical value $t_{p}\sim U/4$, the PSS phase appears in between the PSF and DW(2,2,0,0) solid. As seen from Fig.\ref{Fig6}(c), the PSS phase can transform to the ASS phase with increasing single particle hopping strength $t$. 
In Fig.\ref{Fig6}, the transformation of the phase diagram of the extended Bose-Hubbard model with NNN interaction is illustrated as the paired phases PSF and PSS appear successively due to the increasing pair hopping amplitude $t_{p}$. Since the PSS phase arises as a result of the interplay between long range interaction and pair hopping, which can be engineered \cite{pair_tunneling_zhou}, estimating the appropriate parameter regime is important for emulating this model in ultracold atomic setups for the possible detection of these exotic paired phases in the near future experiments. Moreover, the inclusion of the pair hopping term leads to the intriguing behavior of the transitions, with the appearance of tri-critical points, indicating the possibility of observing these phases as metastable states, which can be accessed experimentally via thermal quench protocols \cite{Ferlaino_SS_quench_expt1,Ferlaino_SS_quench_expt2,
Ferlaino_SS_quench_2d_expt}.

\begin{figure}
	\centering
	\includegraphics[width=1.02\columnwidth]{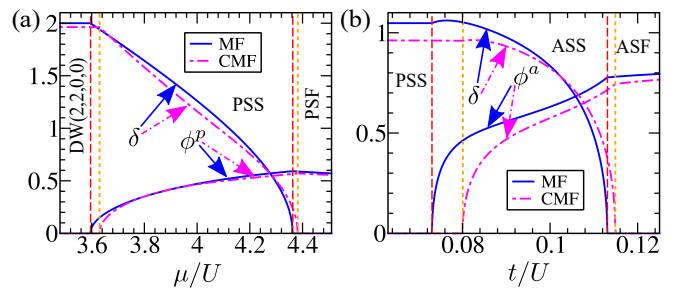}
	\caption{Variation of different order parameters using MF (blue solid lines) and CMF ($2 \times 2$) technique (pink dashed-dotted lines) for fixed pair hopping amplitude $t_{p}/U=0.35$ and NNN interaction strength $V_{2}/U=0.5$: (a) Density imbalance $\delta$ and PSF order $\phi^{p}$ with increasing $\mu$ for $t/U = 0.05$; (b) Density imbalance $\delta$ and ASF order $\phi^{a}$ with increasing $t$ for $\mu/U=4.1$. The vertical red dashed (orange dotted lines) separate the different phases obtained from MF (CMF) theory.}      
\label{Fig9}
\end{figure}

In order to investigate the stability of the PSS phase in presence of fluctuations, we also obtain the phase diagram using the CMF approach, as shown in Fig.\ref{Fig8}. Although the qualitative behavior of the phase diagram remains same as that obtained from single-site MF theory, the phase boundaries (particularly the PSS-ASS boundary) shift due to the inter-site correlations within the cluster. Consequently, the region of ASS phase shrinks in the phase diagram.  As discussed before, at the MF level, the PSS-PSF as well as PSS-DW phase boundaries remain unaffected by the single particle hopping $t$, which now in the CMF theory acquire a weak $t$ dependence.

To characterize various phases and capture the transitions between them, we compute the different order parameters using MF theory, which are also compared with that of the CMF approach. In the PSS phase, apart from finite PSF order $\phi^{p}$, the non-vanishing density difference $\delta$  between the sub-lattices signifies the density ordering, however the atomic SF order parameter $\phi^{a}$ vanishes. As seen from the phase diagram in Fig.\ref{Fig6}(c) and Fig.\ref{Fig8}, changing $\mu$ can lead to melting of DW(2,2,0,0) solid to PSS phase, which finally transforms to the PSF phase. Such transition can be captured from the variation of order parameters $\delta$ and $\phi^{p}$, which is shown in Fig.\ref{Fig9}(a). Moreover, in Fig.\ref{Fig9}(b), the variation of order parameters $\delta$ and $\phi^{a}$ across the PSS-ASS-ASF transition is depicted by tuning the single particle hopping strength $t$.

\subsection{Collective excitations}
To unveil the characteristic features of the PSS phase and its transitions, we next analyze the low-lying collective excitations.

\begin{figure}[b]
	\centering
	\includegraphics[width=1.02\columnwidth]{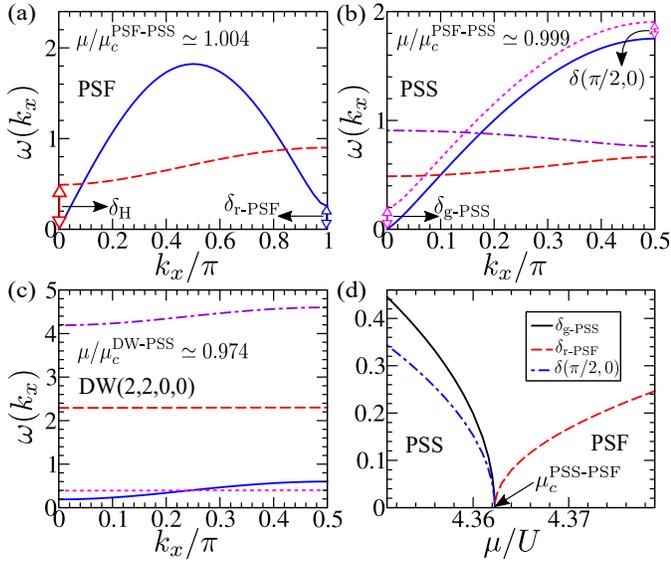}
	\caption{Continuous PSF-PSS-DW transition with decreasing chemical potential $\mu$, for $t/U=0.05$ and $V_{2}/U=0.5$: low-lying collective excitations $\omega(k_{x})$ of the (a) PSF, (b) PSS phase close to the critical point of the PSF-PSS transition, and also for the (c) DW(2,2,0,0) solid. The excitation spectrum in (b,c) are folded within the reduced Brillouin zone due to lattice translational symmetry breaking along the $x$-direction. (d) Variation of different energy gaps with increasing $\mu$ across the PSF-PSS transition. The critical point for PSF to PSS transition is $\mu^{\text{PSF-PSS}}_{c}/U\simeq 4.362$ and for PSS to DW transition is $\mu^{\text{PSS-DW}}_{c}/U\simeq 3.595$.}      
\label{Fig10}
\end{figure}
 
In order to capture the physical picture behind the formation of PSS phase, we first analyze the system in the atomic limit $t \sim 0$. The pair particle/hole excitations of DW(2,2,0,0) solid can lead to its instability towards the formation of the PSS phase, however they cost higher energy compared to the single particle/hole excitations in the atomic limit. Due to pair hopping, such localized pair particle/hole excitations de-localize, resulting in a reduction of their energy which can be calculated within the degenerate perturbation theory, $\epsilon^{2p}=-2\mu+U+16V_{2}-4t_{p}$ and $\epsilon^{2h}=2\mu-U-4t_{p}$. When $\epsilon^{2p/2h}=0$, the instability of DW(2,2,0,0) solid sets in, which yields its boundary at $\mu^{2p}_{c} = U/2+8V_{2}-2t_{p}$ and $\mu^{2h}_{c}=U/2+2t_{p}$. For $t_{p}\gtrsim U/4$, the single particle/hole excitations remain non-vanishing, while the pair particle/hole excitations become unstable at the above-mentioned boundaries, leading to the formation of the PSS phase.
In general, the DW(2,2,0,0) phase has four gapped single and pair particle/hole excitations, the dispersion of which are depicted in Fig.\ref{Fig10}(c). At the MF level, the pair particle/hole excitations are independent of single particle hopping strength $t$ and become gapless at the boundaries of the continuous transition to PSS phase.

\begin{figure}
	\centering
	\includegraphics[width=1.02\columnwidth]{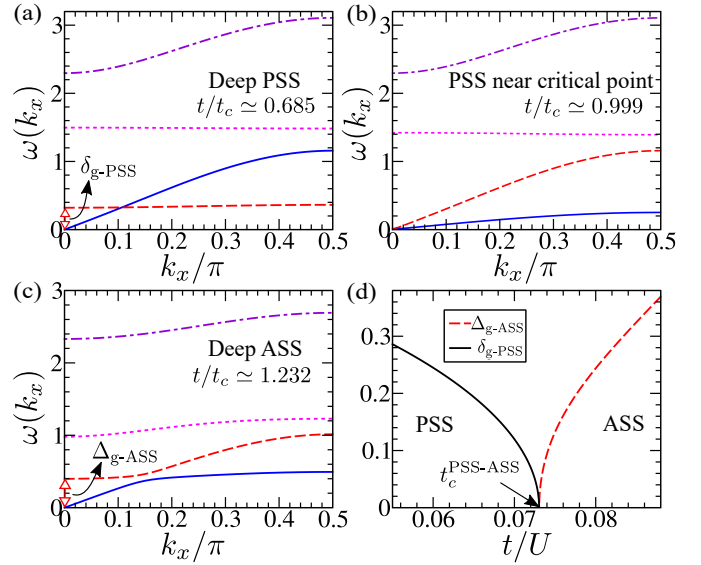}
	\caption{Continuous PSS-ASS transition with increasing single particle hopping strength $t$, for $\mu/U=4.1$ and $V_{2}/U=0.5$: low-energy collective excitations $\omega(k_{x})$ for the (a) deep PSS, (b) PSS phase close to the critical point, as well as for the (c) deep ASS phase. The excitation spectrum are folded within the reduced Brillouin zone due to lattice translational symmetry breaking along the $x$-direction. (d) Variation of the energy gaps of the low-energy gapped mode, $\delta_{\text{g-PSS}}$ (for PSS) and $\Delta_{\text{g-ASS}}$ (for ASS) with increasing $t$ across the PSS-ASS transition point $t_{c}/U\simeq 0.073$ .}      
\label{Fig11}
\end{figure}

As discussed in the previous section, the PSF phase has a gapless sound mode as well as a gapped excitation. However, in the presence of NNN interaction, the sound mode develops a roton minima as a precursor to the formation of PSS phase [see Fig.\ref{Fig10}(a)]. In this case, the softening of the roton mode at $(k_{x},k_{y})=(\pi,0)$ and $(0,\pi)$ indicates a possible stripe density order. As a result of pair superfluidity, a gapless sound mode also appears in the PSS phase. Due to the formation of the stripes in PSS, the lattice translation symmetry is broken and consequently, the excitation spectrum is folded within the reduced Brillouin zone, as depicted in Fig.\ref{Fig10}(b).
Due to such folding in the translational symmetry broken PSS phase, an additional low-energy gapped excitation appears as a reminiscent of the roton mode. The gap of this low-energy excitation ($\delta_{\rm g-PSS}$) in the PSS phase as well as the roton gap ($\delta_{\rm r-PSF}$) in the PSF phase vanishes at the phase boundary of the continuous PSS-PSF transition, as shown in Fig.\ref{Fig10}(d). Due to the translational symmetry breaking, another gap appears at $(\pi/2,0)$ in the lowest-energy excitation, which also vanishes at the boundary of PSS-PSF transition.

Next, we focus on the characteristic changes in the excitation spectrum for the PSS-ASS transition by increasing the single particle hopping strength $t$, which is depicted in Fig.\ref{Fig11}. Since the lattice translational symmetry is broken in both the phases, the excitation spectrum is folded within the reduced Brillouin zone [see Fig.\ref{Fig11}(a-c)]. Similar to the PSF-ASF transition discussed in the previous section, the lowest gapped excitations of both the phases become gapless at the phase boundary of the continuous PSS-ASS transition, as shown in Fig.\ref{Fig11}(d). Note that, for the hole dominated regime [see the phases below the DW(2,2,0,0) phase boundary in Fig.\ref{Fig6}(c)], such gaps do not vanish when the transition becomes first-order.

Above analysis reveals that, similar to the atomic SF, the roton mode softening phenomena as a precursor of the density ordering and appearance of an additional low-energy gapped mode are robust characteristics of supersolidity, which are also present for bosonic pairs.

\section{Melting of pair superfluid and pair supersolid at finite temperatures}
\label{finite_temperature_phase_diagram}
In this section, we investigate the melting of paired states of bosons in the presence of thermal fluctuations and obtain the corresponding phase diagrams at finite temperatures.  To systematically incorporate the thermal and quantum fluctuations, we employ the CMF method at finite temperatures as outlined in Sec.\ref{CMF_theory}, and compare the results with that obtained from the single-site MF theory. 

\begin{figure}[b]
	\centering
	\includegraphics[width=1.02\columnwidth]{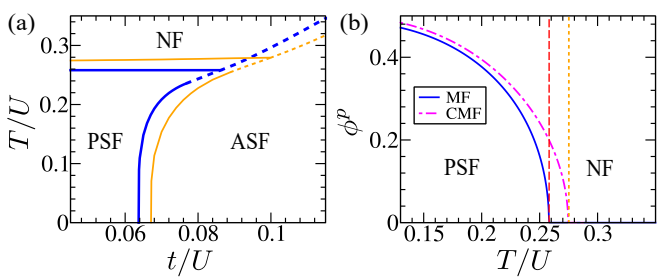}
	\caption{Melting of the PSF phase with increasing temperature $T$, for $t_{p}/U=0.26$ and $\mu/U=-0.25$: (a) Comparison between MF (thick lines) and CMF-$2\times2$ (thin lines) phase diagram in the $T/U$-$t/U$ plane. The thick and thin solid (dotted) lines correspond to the continuous (first-order) transition under MF and CMF theory, respectively. (b) Variation of the PSF order $\phi^{p}$ with increasing temperature across the continuous PSF-NF transition for fixed value of single particle hopping strength $t/U=0.05$. The blue solid and pink dashed-dotted lines are obtained within MF and CMF ($2 \times 2$) approach, respectively, where the vertical red dashed (orange dotted) lines indicate the critical point obtained from the MF (CMF) theory.}      
\label{Fig12}
\end{figure}

With increasing the temperature, all the phases eventually melt to the homogeneous normal fluid (NF), where all the order parameters vanish. First, we focus on the melting of the PSF phase in the absence of NNN interaction ($V_{2}=0$), and obtain the phase diagram in the $T/U$-$t/U$ plane, as shown in Fig.\ref{Fig12}(a). It is evident from the phase diagram, two first-order boundaries meet with a line of continuous transition from PSF to NF, at a particular point in the $T/U$-$t/U$ plane. The PSF order $\phi^{p}$ gradually decreases with temperature and vanishes in the NF phase [see Fig.\ref{Fig12}(b)]. Note that, the nature of such transition can be first-order or continuous, which crucially depends on the physical parameters, particularly the pair hopping strength $t_{p}$.
As expected, the critical temperature for ASF-NF transition scales linearly with the single particle hopping $t$, however on the contrary, the melting temperature of PSF to NF is almost independent of $t$. Due to the pair hopping, the ASF-NF  and ASF-PSF transition can be of first-order type [as seen from Fig.\ref{Fig12}(a)], indicating the  existence of metastable (pair) superfluid phase at finite temperatures, which can be observed by quenching the temperature from NF.

\begin{figure}
	\centering
	\includegraphics[width=1.02\columnwidth]{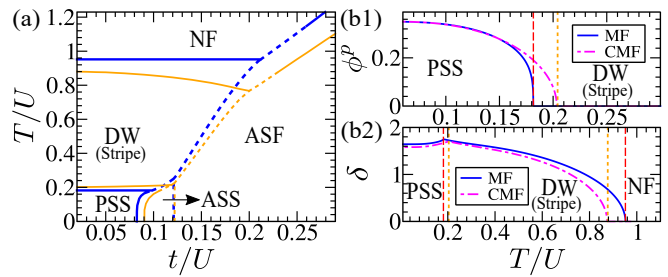}
	\caption{Melting of the PSS phase with increasing temperature $T$, for $t_{p}/U=0.35$, $\mu/U=3.8$, and $V_{2}/U = 0.5$: (a) Comparison between MF (thick lines) and CMF-$2\times2$ (thin lines) phase diagram in the $T/U$-$t/U$ plane. The thick and thin solid (dotted) lines correspond to the continuous (first-order) transition under MF and CMF theory, respectively. Variation of the (b1) PSF order $\phi^{p}$ and (b2) density order $\delta$ with increasing temperature across the continuous PSF-DW-NF transition for a fixed value of atomic hopping strength $t/U=0.05$. The blue solid and pink dashed-dotted lines are obtained within MF and CMF ($2 \times 2$) approach, respectively, where the vertical red dashed (orange dotted) lines separate the different phases obtained within the MF (CMF) theory.}      
\label{Fig13}
\end{figure}

Next, we study the melting of the PSS phase which is formed for sufficiently strong NNN interaction strength $V_{2}$. 
The phase diagram in the $T/U$-$t/U$ plane is obtained to illustrate the region of stable PSS phase at finite temperature and its melting process [see Fig.\ref{Fig13}(a)]. In the two step melting process of the PSS phase, first the PSF order $\phi^{p}$ disappears with increasing the temperature, leading to the formation of stripe density wave (DW), which finally  melts to form the homogeneous NF, as shown in Fig.\ref{Fig13}(b1,b2). Similar to the melting temperature of the PSF phase, the PSS-DW transition temperature remains almost independent of single particle hopping strength $t$. As evident from the phase diagram, the first-order boundary between stripe DW to ASF connects the two special points, where three phase boundaries meet, resulting in a competition between the different orders. Interestingly, except for the appearance of the PSS phase at low temperatures, the phase diagram in Fig.\ref{Fig13}(a) exhibits similarity with that of the hardcore bosons in absence of pair hopping \cite{Manali_CMFT_2}. 

Although the qualitative features of the phase diagram remain same, the inclusion of correlations in CMF approach enhances the region of stability of the paired phases (PSF and PSS) in the  $T/U$-$t/U$ plane, with an increase in their corresponding melting temperatures. On the contrary, the cluster correlations reduce the transition temperature of stripe DW to NF phase. 
From the comparison between Fig.\ref{Fig12}(a) and \ref{Fig13}(a), we notice that, in presence of the NNN interaction, the inclusion of cluster correlations yields a significant correction over MF theory at higher temperatures, which is reflected from the shift in the phase boundaries.

The qualitative picture for the formation of the stable paired phases at finite temperatures is important for their detection, since the thermal fluctuations are inevitable in experiments. Moreover, the stable and metastable paired phases can be achieved by quenching the temperature, as done in  recent experiments to obtain the supersolid state of dipolar gas \cite{Ferlaino_SS_quench_expt1, Ferlaino_SS_quench_expt2,Ferlaino_SS_quench_2d_expt}.

\section{conclusion}
\label{conclusion}
To summarize, we have investigated in details the formation of pair superfluid (PSF) and pair supersolid (PSS) phases arising due to the presence of engineered pair hopping as well as NNN interaction mimicking the effect of long-range forces. The characteristic properties of such exotic phases of bosons and nature of their transitions are also analyzed from the excitation spectrum. The effect of correlations are systematically taken into account both at zero as well as finite temperatures within the scheme of cluster mean field (CMF) theory. We consider bosons in the presence of three-body constraint, which in turn enhances the stability of PSF phase as well as enables us to obtain the analytical results. 
Above a critical pair hopping strength, the pair particle/hole production in the insulating MI-0 and MI-2 phases leads to the formation of PSF phase, where the atomic SF order vanishes.
As a consequence of pair SF order, the PSF phase exhibits a gapless sound mode for which the sound velocity scales with pair hopping strength. Similar to the Higgs mode in ASF, a gapped mode also appears in PSF phase, the gap of which decreases with the single particle hopping strength as a result of pair breaking. The competition between single particle and pair hopping leads to the PSF to ASF transition exhibiting interesting behavior due to the presence of a tri-critical point, where the nature of transition changes from first order to continuous. Moreover, we also identify the region where PSF phase can exist as a metastable state. For continuous PSF-ASF transition, the energy gap of first excitation of both the phases vanishes at the critical point, which occurs for Ising class transitions.  The present model has $U(1) \times \mathbb{Z}_{2}$ symmetry \cite{PSF_square_lattice_QMC_Wessel,PSS_triangular_lattice_Zhang,Christophe_Mora_PRL}, where the $U(1)$ symmetry broken superfluid phases undergo a PSF to ASF transition, breaking the remaining $\mathbb{Z}_{2}$ symmetry, which falls under the Ising class \cite{PSF_square_lattice_QMC_Wessel}. 
 
For sufficiently strong NNN interaction strength, the homogeneous Mott-1 (MI-1) phase is replaced by a stripe ordered solid of pair particles, which is crucial for the formation of PSS phase. 
Such pair solid melts above a critical pair hopping strength, due to the pair particle/hole production, giving rise to the PSS phase with coexisting pair superfluidity and stripe density order. Even for such paired states of bosons, the roton mode softening as a precursor of density ordering and appearance of a low-energy gapped mode are robust signatures of supersolidity \cite{Shlyapnikov_roton,Nozieres_roton,amplitude_mode_SS,
Higgs_mode_SS_expt,Manali_CMFT_2}, which can be probed in the experiments. In this context, for the formation of atomic supersolid, both the roton as well as gapped Higgs mode are detected in the recent experiments \cite{Ferlaino_expt1_roton,Ferlaino_expt2_roton,Ferlaino_expt3_roton,
TPfau_expt1_roton,TPfau_expt2_roton,Esslinger_expt1_roton,
Higgs_mode_SS_expt}. The recent proposal to engineer the pair hopping using external laser fields seems to be promising for realizing such unconventional paired states of bosons \cite{pair_tunneling_zhou}. Also, the attractive interaction between the bosons can effectively generate the pair hopping \cite{Lewenstein_PSS}, which is a key ingredient for the formation of such paired states, as discussed in the present work. Moreover, we believe our analysis can also shed light on the pairing between two component bosons. Due to the presence of long-range interactions, engineering the pairing between the dipolar atoms \cite{Lewenstein_pair_hopping,Lewenstein_PSS} can open up the possibility to explore the exotic pair supersolid state, which can also be important for understanding superconductivity.

In order to investigate the stability of such paired states under thermal fluctuations, the finite temperature phase diagrams are obtained using the CMF theory. We also focus on the melting pathway of pair supersolid phase, which is particularly interesting due to the coexisting pair superfluidity and density order. 
In contrast to the atomic phases, the inclusion of cluster correlations in CMF theory yields a larger value of the melting temperature of the paired phases compared to that of MF theory, indicating the stability of these phases at finite temperatures.
Moreover, as a result of pair hopping, the transition from ASF to normal fluid (NF) can become first-order. Near the first-order transition, the metastable configurations of the paired phases can be explored by a temperature quench, as done in recent experiments on dipolar supersolids \cite{Ferlaino_SS_quench_expt1, Ferlaino_SS_quench_expt2,Ferlaino_SS_quench_2d_expt}.
The interesting results discussed in this work are based on the framework of mean field and cluster mean field approach, which deserve further attention by using more accurate techniques like Quantum Monte Carlo.

The present analysis elucidates the basic phenomenology behind the formation of such exotic paired states of bosons as well as the robust features captured from their low-lying excitations, which are of relevance for the ongoing cold atom experiments.

\appendix
\section{Pair superfluid phase with varying pair hopping strength}
\label{app1}
\begin{figure}
	\centering
	\includegraphics[width=1.02\columnwidth]{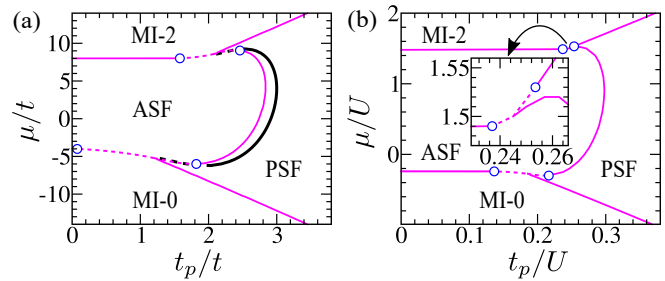}
	\caption{Zero temperature phase diagrams for varying pair hopping strength $t_{p}$: in the (a) $\mu/t$-$t_{p}/t$ and (b) $\mu/U$-$t_{p}/U$ plane. The thin lines are obtained using $2 \times 2$ clusters, whereas the thick lines in (a) are obtained using MF theory. Solid (dotted) lines correspond to the continuous (first-order) transition. The inset in (b) shows the zoomed area of the phase diagram pointed by the curved arrow. The circular symbols denote the tri-critical points obtained within the CMF framework, where the first order and continuous phase boundaries meet. Parameters chosen: (a) $U$, $V_{2}=0$, and (b) $t/U = 0.06$, $V_{2}=0$.}      
\label{Fig14}
\end{figure}
Here, we study the formation of pair superfluid phase (PSF) with varying pair hopping amplitude $t_{p}$. To do so, we compute the zero temperature phase diagram in the $\mu/t$-$t_{p}/t$ as well as $\mu/U$-$t_{p}/U$ planes within the CMF approach [see Fig.\ref{Fig14}]. Interestingly, in the absence of on-site and NNN interactions ($U,V_{2}=0$), it is observed that the phase diagram in Fig.\ref{Fig14}(a) appears qualitatively similar to the QMC results obtained for triangular lattice [see Fig.2(a) of Ref.\cite{PSS_triangular_lattice_Zhang}].  By  incorporating the correlations using CMF theory, the ASF-PSF phase boundary in Fig.\ref{Fig14}(a) shifts and the region of PSF phase increases, indicating its enhanced stability. However, the nature of this transition differs from the QMC result of Ref.\cite{PSS_triangular_lattice_Zhang}, where the ASF-PSF transition turns out to be first-order. On the contrary, within the CMF approach, the ASF-PSF boundary correspond to continuous transition, which changes it nature with the appearance of two tri-critical points, as shown in Fig.\ref{Fig14}(a). Here we point out that, the first order jump across the ASF-PSF transition (as shown in Ref.\cite{PSS_triangular_lattice_Zhang}) can only be captured using the finite size scaling analysis, which is beyond the scope of the present calculations performed for finite size clusters. Although the accuracy of the phase boundaries can be improved using CMF theory, however in some cases it fails to capture the exact nature of the transition, particularly in two dimensions.

We also compute the phase diagram in $\mu/U$-$t_{p}/U$ plane using CMF approach, as shown in Fig.\ref{Fig14}(b). The qualitative behavior as well as the nature of transition turns out to be similar to that obtained for triangular lattice  using MF theory, as seen from Fig.6(b) of Ref.\cite{CPB_Zhang}.

\section{Pair superfluid phase for attractive interactions}
\label{app2} 
\begin{figure}[b]
	\centering
	\includegraphics[width=0.85\columnwidth]{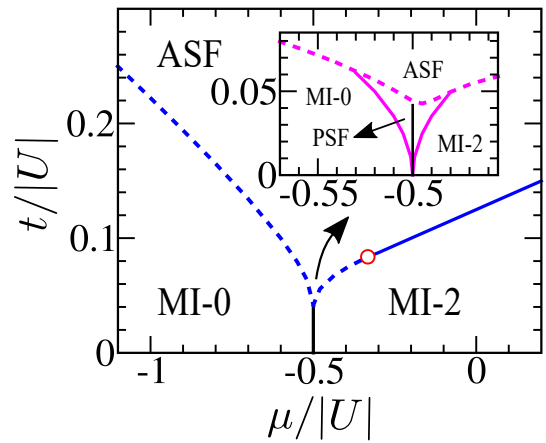}
	\caption{Zero temperature mean field phase diagram in the presence of attractive on-site interaction $U<0$ for $t_{p}$, $V_{2}=0$. The inset shows the mean field phase diagram for $t_{p}\sim 2t^2/|U|$ where the PSF phase appears [see the text for details]. Solid black vertical line denotes the boundary between the MI-0 and MI-2 phases for $t_{p}=0$. The solid (dotted) lines denote continuous (first-order) transitions and the circular symbol denotes the tri-critical point.}      
\label{Fig15}
\end{figure}

To this end, we study the phase diagram in the presence of attractive on-site interaction ($U<0$) under three body constraint, and investigate the PSF phase within the mean field (MF) analysis. In the presence of attractive interaction, previous Quantum Monte Carlo (QMC) studies have reported the formation of PSF phase, where the quantum fluctuations play an important role \cite{PSF_square_lattice_QMC_Wessel,PSF_square_lattice_QMC_K_Ng}.  Here, we attempt to explain the appearance of such paired phase within the framework of simple MF analysis. The phase diagram of attractive Bose-Hubbard model with $U<0$ and $t_{p}=0$ in the presence of three body constraint is illustrated in Fig.\ref{Fig15}. As evident from the phase diagram, the MF theory can only capture the Mott and ASF phases, whereas the PSF phase does not appear in absence of the pair hopping term $t_{p}$. However, a careful analysis across the boundary between MI-0 and MI-2 phase at $\mu=-|U|/2$ shows that, at the MF level, any linear combination of the $\ket{0}$ and $\ket{2}$ state at any site i.e $\ket{\psi}=\cos{\theta}\ket{0}+\sin{\theta}\ket{2}$ has zero energy. The inclusion of an infinitesimally small pair hopping leads to an additional energy $-zt_{p}\sin^2{2\theta}/2$, which is minimized for $\theta=\pi/4$. This state corresponds to a PSF phase along the line at $\mu=-|U|/2$, separating the MI-0 and MI-2 phases. This argument justifies the appearance of the PSF phase in the limit $t_{p}\rightarrow 0^{+}$  across the above mentioned line. From the excitation spectrum in Eq.~\eqref{psf_excitations} with $U=-|U|$, it can be seen that the PSF phase has a zero energy Goldstone mode $\omega^{\rm G}=0$ and a gapped mode with energy $\delta_{\rm H}=\sqrt{4t^2-6t|U|+|U|^2/4}$, which vanishes at the end point of the border between the MI-0 and MI-2 phases. 

As seen from the above analysis, the PSF phase can also be obtained in an extended region of the phase diagram even from the MF analysis if the pair hopping term is included in the effective Hamiltonian. By using the perturbative analysis outlined in Ref.\cite{Lewenstein_PSS}, the single particle hopping term can generate an effective pair hopping with amplitude $t_{p}\sim 2t^2/|U|$ for $t\ll |U|$. The inclusion of such pair hopping term broadens the region of the PSF phase around the line $\mu = -|U|/2$ in between two Mott phases, as shown in the inset of Fig.\ref{Fig15}. Interestingly, the MF analysis of the effective Hamiltonian including the fluctuation generated pair hopping term qualitatively reproduces the phase diagram obtained using QMC analysis of Ref.\cite{PSF_square_lattice_QMC_Wessel}. We can also calculate the excitation spectrum of different phases using the equations given in Sec.\ref{collective_excitations_PSF} with appropriate parameters as mentioned above. For consistency, we have also checked that the vanishing of the excitations of MI-0 and MI-2 phases [from Eq.~\eqref{MI-0_p_branch} and Eq.~\eqref{MI-2_p_branch}] yields the phase boundaries $\mu = -|U|/2\mp 8t^2/|U|$ between PSF and Mott phases with continuous transition.

\begin{thebibliography}{99}

\bibitem{Sanpera_review}
M. Lewenstein, A. Sanpera, V. Ahufinger, B. Damski, A. Sen, and U. Sen, Adv. Phys. {\bf 56}, 243 (2007).

\bibitem{Dalibard_review}
I. Bloch, J. Dalibard, and W. Zwerger, Rev. Mod. Phys. {\bf 80}, 885 (2008).


\bibitem{Modugno_SS_expt1}
L. Tanzi, E. Lucioni, F. Fam\`{a}, J. Catani, A. Fioretti, C. Gabbanini, R. N. Bisset, L. Santos, and G. Modugno, Phys. Rev. Lett.
{\bf 122}, 130405 (2019).

\bibitem{TPfau_SS_expt1}
 F. B\"{o}ttcher, J. Schmidt, M. Wenzel, J. Hertkorn, M. Guo, T. Langen, and T. Pfau, Phys. Rev. X {\bf 9}, 011051 (2019).


\bibitem{Ferlaino_SS_quench_expt1} 
L. Chomaz, D. Petter, P. Ilzh\"{o}fer, G. Natale, A. Trautmann, C. Politi, G. Durastante, R. M. W. van Bijnen, A. Patscheider, M. Sohmen, M. J. Mark, F. Ferlaino, Phys. Rev. X {\bf 9}, 021012 (2019).

\bibitem{Ferlaino_SS_quench_expt2}
 M. Sohmen, C. Politi, L. Klaus, L. Chomaz, M. J. Mark, M. A. Norcia, and F. Ferlaino, Phys. Rev. Lett. {\bf 126}, 233401 (2021).


\bibitem{Ferlaino_SS_2d_expt1}
M. A. Norcia, C. Politi, L. Klaus, E. Poli, M. Sohmen, M. J. Mark, R. N. Bisset, L. Santos, and F. Ferlaino, Nature (London) {\bf 596}, 357 (2021).


\bibitem{Ferlaino_SS_quench_2d_expt}
T. Bland, E. Poli, C. Politi, L. Klaus, M. A. Norcia, F. Ferlaino, L. Santos, R. N. Bisset, Phys. Rev. Lett. {\bf 128}, 195302 (2022).



\bibitem{Penrose_BEC}
O. Penrose and L. Onsager, Phys. Rev. {\bf 104}, 576 (1956).

\bibitem{Lifshitz}
A. F. Andreev and I. M. Lifshitz, Sov. Phys. JETP {\bf 29}, 1107 (1969).

\bibitem{Chester}
G. V. Chester, Phys. Rev. A {\bf 2}, 256 (1970).

\bibitem{Leggett}
 A. J. Leggett, Phys. Rev. Lett. {\bf 25}, 1543 (1970).
 
\bibitem{Fisher} 
K.-S. Liu and M. E. Fisher, J. Low Temp. Phys. {\bf 10}, 655 (1973).
 
\bibitem{supersolids_review}
M. Boninsegni and N. V. Prokof'ev, Rev. Mod. Phys. {\bf 84}, 759 (2012). 


\bibitem{Daley_repulsive_bound_pairs}
K. Winkler, G. Thalhammer, F. Lang, R. Grimm, J. H. Denschlag, A. J. Daley, A. Kantian, H. P. B\"{u}chler, and P. Zoller, Nature (London) {\bf 441}, 853 (2006).

\bibitem{Bloch_second_order_tunneling}
S. F\"{o}lling, S. Trotzky, P. Cheinet, M. Feld, R. Saers, A.
Widera, T. M\"{u}ller, and I. Bloch, Nature (London) {\bf 448}, 1029 (2007).

\bibitem{Lukin_photon_bound_states}
O. Firstenberg, T. Peyronel, Q.-Y. Liang, A. V. Gorshkov, M. D. Lukin, and V. Vuleti\'{c}, Nature (London) {\bf 502}, 71 (2013).

\bibitem{Rippol_PRA}
M. Eckholt and J. J. Garc\'{i}a-Ripoll, Phys. Rev. A {\bf 77}, 063603 (2008).


\bibitem{PSF_square_lattice_QMC_Wessel}
L. Bonnes and S. Wessel, Phys. Rev. Lett. {\bf 106}, 185302 (2011).

\bibitem{PSF_square_lattice_QMC_K_Ng}
Y.-C. Chen, K.-K. Ng, and M.-F. Yang, Phys. Rev. B {\bf 84}, 092503 (2011).


\bibitem{Lewenstein_PRA}
T. Sowi\'{n}ski, R. W. Chhajlany, O. Dutta, L. Tagliacozzo, and M. Lewenstein, Phys. Rev. A {\bf 92}, 043615 (2015).

\bibitem{Yang_PRA}
Y.-W. Lee and M.-F. Yang, Phys. Rev. A {\bf 81}, 061604(R) (2010).

\bibitem{Daley_PRL}
A. J. Daley, J. M. Taylor, S. Diehl, M. Baranov, and P. Zoller, Phys. Rev. Lett. {\bf 102}, 040402 (2009).



\bibitem{two_component_Kuklov_PRL}
A. Kuklov, N. Prokof’ev, and B. Svistunov, Phys. Rev. Lett. {\bf 92}, 050402 (2004).

\bibitem{two_component_Mathey_PRA}
A. Hu, L. Mathey, I. Danshita, E. Tiesinga, C. J. Williams, and C. W. Clark, Phys. Rev. A {\bf 80}, 023619 (2009).

\bibitem{two_component_Stringari_PRA}
C. Menotti and S. Stringari, Phys. Rev. A {\bf 81}, 045604 (2010).

\bibitem{PSF_Quantum_Gutzwiller}
V. E. Colussi, F. Caleffi, C. Menotti, and A. Recati, SciPost Phys. {\bf 12}, 111 (2022).

\bibitem{two_layer_Luis} 
A. Arg\"{u}elles and L. Santos, Phys. Rev. A {\bf 75}, 053613 (2007).

\bibitem{PSF_Bradraj} 
B. Pandey, S. Sinha, and S. K. Pati, Phys. Rev. B {\bf 91}, 214432 (2015).

\bibitem{pair_tunneling_zhou}
 X. F. Zhou, Y. S. Zhang, and G. C. Guo, Phys. Rev. A {\bf 80}, 013605 (2009).


\bibitem{Christophe_Mora_PRL}
J. Lebreuilly, C. Aron, and C. Mora, Phys. Rev. Lett. {\bf 122}, 120402 (2019).

\bibitem{Tapan_Mishra_1} 
M. Singh, S. Greschner, and T. Mishra, Phys. Rev. A {\bf 98}, 023615 (2018).

\bibitem{Tapan_Mishra_2} 
S. Lahiri, S. Mondal, K. Pandey, and T. Mishra, Phys. Rev. A {\bf 102}, 043710 (2020).


\bibitem{Lewenstein_pair_hopping}
T. Sowi\'{n}ski, O. Dutta, P. Hauke, L. Tagliacozzo, and M. Lewenstein, Phys. Rev. Lett. {\bf 108}, 115301 (2012).

\bibitem{Sandvik_pair_hopping_QMC} 
A. J. R. Heng, W. Guo, A. W. Sandvik, and P. Sengupta, Phys. Rev. B {\bf 100}, 104433 (2019).

\bibitem{Sebastian_jc1}
L. Guo, S. Greschner, S. Zhu, and W. Zhang, Phys. Rev. A {\bf 100}, 033614 (2019).


\bibitem{Lewenstein_PSS}
C. Trefzger, C. Menotti, and M. Lewenstein, Phys. Rev. Lett. {\bf 103}, 035304 (2009).

\bibitem{PSS_triangular_lattice_Zhang}
W. Zhang, R. Yin, and Y. Wang, Phys. Rev. B {\bf 88}, 174515 (2013).

\bibitem{CPB_Zhang} 
Y. C. Wang, W. Z. Zhang, H. Shao, and W. A. Guo, Chin. Phys. B 22, 096702 (2013).

\bibitem{Shlyapnikov_roton}
L. Santos, G. V. Shlyapnikov, and M. Lewenstein, Phys. Rev. Lett. {\bf 90}, 250403 (2003).

\bibitem{Nozieres_roton}
P. Nozi\`{e}res, J. Low Temp. Phys. {\bf 137}, 45 (2004).


\bibitem{Ferlaino_expt1_roton}
L. Chomaz, R. M. W. van Bijnen, D. Petter, G. Faraoni, S. Baier, J. H. Becher, M. J. Mark, F. W\"{a}chtler, L. Santos, and F. Ferlaino, Nat. Phys. {\bf 14}, 442 (2018).

\bibitem{Ferlaino_expt2_roton}
D. Petter, G. Natale, R. M. W. van Bijnen, A. Patscheider, M. J. Mark, L. Chomaz, and F. Ferlaino, Phys. Rev. Lett. {\bf 122}, 183401 (2019).

\bibitem{Ferlaino_expt3_roton}
G. Natale, R. M. W. van Bijnen, A. Patscheider, D. Petter, M. J. Mark, L. Chomaz, and F. Ferlaino, Phys. Rev. Lett. {\bf 123}, 050402 (2019).

\bibitem{TPfau_expt1_roton}
J. N. Schmidt, J. Hertkorn, M. Guo, F. B\"{o}ttcher, M. Schmidt, K. S. H. Ng, S. D. Graham, T. Langen, M. Zwierlein, and T. Pfau, Phys. Rev. Lett. {\bf 126}, 193002 (2021).

\bibitem{TPfau_expt2_roton}
J. Hertkorn, J. N. Schmidt, F. B\"{o}ttcher, M. Guo, M. Schmidt, K. S. H. Ng, S. D. Graham, H. P. B\"{u}chler, T. Langen, M. Zwierlein, and T. Pfau, Phys. Rev. X {\bf 11}, 011037 (2021).


\bibitem{Esslinger_expt1_roton}
R. Mottl, F. Brennecke, K. Baumann, R. Landig, T. Donner, and T. Esslinger, Science 336, 1570 (2012).

\bibitem{K_Ng_QMC_PSF}
K.-K. Ng and M.-F. Yang, Phys. Rev. B {\bf 83}, 100511(R) (2011).



\bibitem{Yamamoto_triangular_CMFT}
D. Yamamoto, I. Danshita, and C. A. R. S\'{a} de Melo, Phys. Rev.
A {\bf 85}, 021601(R) (2012).

\bibitem{Yamamoto_square_CMFT}
D. Yamamoto, A. Masaki, and I. Danshita, Phys. Rev. B {\bf 86}, 054516 (2012).

\bibitem{Heydarinasab_CMFT}
F. Heydarinasab and J. Abouie, Phys. Rev. B {\bf 96}, 104406 (2017).

\bibitem{Demler_CMFT}
D. Pekker, B. Wunsch, T. Kitagawa, E. Manousakis, A. S. S\o{}rensen, and E. Demler, Phys. Rev. B 86, 144527 (2012).

\bibitem{Luhmann_CMFT}
D. S. L\"{u}hmann, Phys. Rev. A {\bf 87}, 043619 (2013).

\bibitem{Fazio_CMFT}
J. Jin, A. Biella, O. Viyuela, L. Mazza, J. Keeling, R. Fazio, and D. Rossini, Phys. Rev. X {bf 6}, 031011 (2016).

\bibitem{Suthar_CMFT_1}
R. Bai, S. Bandyopadhyay, S. Pal, K. Suthar, and D. Angom, Phys. Rev. A {\bf 98}, 023606 (2018).

\bibitem{Suthar_CMFT_2}
K. Suthar, H. Sable, R. Bai, S. Bandyopadhyay, S. Pal, and D. Angom, Phys. Rev. A {\bf 102}, 013320 (2019).

\bibitem{Suthar_CMFT_3}
K. Suthar, R. Kraus, H. Sable, D. Angom, G. Morigi, and J. Zakrzewski, Phys. Rev. B {\bf 102}, 214503 (2020).


\bibitem{Manali_CMFT_1}
M. Malakar, S. Ray, S. Sinha, and D. Angom, Phys. Rev. B {\bf 102}, 184515 (2020).

\bibitem{Sayak_CMFT}
U. Pohl, S. Ray, and J. Kroha, Ann. Phys. (Berlin) {\bf 534}, 2100581 (2022).

\bibitem{Manali_CMFT_2}
M. Malakar, S. Sinha, and S. Sinha, J. Stat. Mech. (2023) 043104.

\bibitem{P_Zoller_original}
D. Jaksch, C. Bruder, J. I. Cirac, C. W. Gardiner, and P. Zoller, Phys. Rev. Lett. {\bf 81}, 3108 (1998).
 
\bibitem{Krauth}
W. Krauth, M. Caffarel, and J. Bouchaud, Phys. Rev. B {\bf 45}, 3137 (1992).

\bibitem{Sinha_EPL}
D. L. Kovrizhin, G. V. Pai, and S. Sinha, Europhys. Lett. {\bf 72}, 162 (2005).

\bibitem{Sinha_unpublished}
D. L. Kovrizhin, G. V. Pai, and S. Sinha, arXiv:0707.2937.


\bibitem{exc_spectra_QMC1}
P. Pippan, H. G. Evertz, and M. Hohenadler, Phys. Rev. A {\bf 80}, 033612 (2009). 

\bibitem{exc_spectra_QMC2}
O. F. Sylju$\mathring{\rm a}$sen, Phys. Rev. B {\bf 78}, 174429 (2008). 

\bibitem{exc_spectra_QMC3}
M. Gartner, F. Mazzanti, and R. E. Zillich, SciPost Phys. {\bf 13}, 025 (2022). 

\bibitem{exc_spectra_Pelster}
M. Ohliger and A. Pelster, World J. Condens. Matter Phys. {\bf 3}, 125 (2013).


\bibitem{exc_spectra_TEBD}
B. Gr\'{e}maud and G. Batrouni, Phys. Rev. B {\bf 93}, 035108 (2016). 

\bibitem{exc_spectra_tMPS}
L. Villa, J. Despres, and L. Sanchez-Palencia, Phys. Rev. A {\bf 100}, 063632 (2019). 

\bibitem{quantum_gutzwiller_Carusotov}
F. Caleffi, M.Capone, C. Menotti, I. Carusotto, and A. Recati, Phys. Rev. Research {\bf 2}, 033276 (2020).

\bibitem{BKT_transition}
V. L. Berezinskii, Sov. Phys. JETP {\bf 32}, 493 (1971); J. M. Kosterlitz and D. J. Thouless, J. Phys. C: Solid State Phys. {\bf 6}, 1181 (1973).

\bibitem{cluster_note}
To do the finite cluster-size scaling, we use the scaling parameter $\lambda=2N_{B}/N_{\mathcal{C}}z$ (introduced in Ref.\cite{Yamamoto_triangular_CMFT}), where $N_{B}$ and $N_{\mathcal{C}}$ denotes the number of bonds and lattice sites in cluster $\mathcal{C}$, respectively, with $z=4$ being the coordination number for square lattice.


\bibitem{Higgs_Altman}
E. Altman and A. Auerbach, Phys. Rev. Lett. {\bf 89}, 250404 (2002); S. D. Huber, E. Altman, H. P. B\"{u}chler, and G. Blatter, Phys. Rev. B {\bf 75}, 085106 (2007).

\bibitem{Higgs_review}
D. Pekker and C. M. Varma, Annu. Rev. Condens. Matter Phys. {\bf 6}, 269 (2015).

\bibitem{Higgs_expt_SF_1}
U. Bissbort, S. G\"{o}tze, Y. Li, J. Heinze, J. S. Krauser, M. Weinberg, C. Becker, K. Sengstock, and W. Hofstetter, Phys. Rev. Lett. {\bf 106}, 205303 (2011).

\bibitem{Higgs_expt_SF_2}
M. Endres, T. Fukuhara, D. Pekker, M. Cheneau, P. Schau\ss, C. Gross, E. Demler, S. Kuhr, and I. Bloch, Nature (London) {\bf 487}, 454 (2012).

\bibitem{Subir_Sachdev}
S. Sachdev, {\it Quantum Phase Transitions} (Cambridge University Press, Cambridge, England, 1999).

\bibitem{Batrouni_QMC_1}
G. G. Batrouni, R. T. Scalettar, G. T. Zimanyi, and A. P. Kampf, Phys. Rev. Lett. {\bf 74}, 2527 (1995); R. T. Scalettar, G. G. Batrouni, A. P. Kampf, and G. T. Zimanyi, Phys. Rev. B {\bf 51}, 8467 (1995).

\bibitem{Batrouni_QMC_2}
G. G. Batrouni and R. T. Scalettar, Phys. Rev. Lett. {\bf 84}, 1599 (2000).

\bibitem{Higgs_mode_SS_expt}
J. L\'{e}onard, A. Morales, P. Zupancic, T. Donner, and T. Esslinger, Science {\bf 358}, 1415 (2017)

\bibitem{amplitude_mode_SS}
J. Hertkorn, F. B\"{o}ttcher, M. Guo, J. N. Schmidt, T. Langen, H. P. B\"{u}chler, and T. Pfau, Phys. Rev. Lett. {\bf 123}, 193002 (2019).


\end{thebibliography}
\end{document}